\documentclass[twocolumn, times, trackchanges]{aastex63_bren}

\usepackage[ruled,vlined]{algorithm2e}
\usepackage{graphicx}
\usepackage{bbm}
\usepackage{float}
\usepackage[hypertexnames=false]{hyperref}
\usepackage{subfiles} 

\usepackage{amsmath}
\hypersetup{linkcolor=red,citecolor=green,filecolor=cyan,urlcolor=magenta}
\shorttitle{multiphased flows in post-starbursts}

\begin{document}

\title{Not so windy after all: MUSE disentangles AGN-driven winds from merger-induced flows in rapidly-transitioning galaxies}

\author{Dalya Baron} 
\email{dalyabaron@gmail.com}
\affiliation{The Observatories of the Carnegie Institution for Science. 813 Santa Barbara Street, Pasadena, CA 91101, USA}

\author{Hagai Netzer} 
\affiliation{School of Physics and Astronomy, Tel-Aviv University, Tel Aviv 69978, Israel}

\author{Dieter Lutz} 
\affiliation{Max-Planck-Institut f$\ddot{u}$r Extraterrestrische Physik, Giessenbachstrasse 1, 85748 Garching, Germany}

\author{Ric I. Davies} 
\affiliation{Max-Planck-Institut f$\ddot{u}$r Extraterrestrische Physik, Giessenbachstrasse 1, 85748 Garching, Germany}

\author{J. Xavier Prochaska} 
\affiliation{Department of Astronomy and Astrophysics, UCO/Lick Observatory, University of California, 1156 High Street, Santa Cruz, CA 95064, USA}



\begin{abstract}

Post-starburst galaxies are believed to be in a rapid transition between major merger starbursts and quiescent ellipticals, where AGN feedback is suggested as one of the processes responsible for the quenching. To study the role of AGN feedback, we constructed a sample of post-starburst candidates with AGN and indications of ionized outflows. We use MUSE/VLT observations to spatially-resolve the properties of the stars and multi-phased gas in five of them. All the galaxies show signatures of interaction/merger in their stellar or gas properties, with some galaxies at an early stage of interaction with companions at distances $\sim$50 kpc, suggesting that optical post-starburst signatures may be present well before the final starburst and coalescence. We detect narrow and broad kinematic components in multiple transitions in all the galaxies. Our detailed analysis of their kinematics and morphology suggests that, contrary to our expectation, the properties of the broad kinematic components are inconsistent with AGN-driven winds in 3 out of 5 galaxies. The two exceptions are also the only galaxies in which spatially-resolved NaID P-Cygni profiles are detected. In some cases, the observations are more consistent with interaction-induced galactic-scale flows, an often overlooked process. These observations raise the question of how to interpret broad kinematic components in interacting and perhaps also in active galaxies, in particular when spatially-resolved observations are not available or cannot rule out merger-induced galactic-scale motions. We suggest that NaID P-Cygni profiles are more effective outflow tracers, and use them to estimate the energy that is carried by the outflow.

\end{abstract}


\section{Introduction}\label{s:intro}

Active galactic nucleus (AGN) feedback is now a key process in models of galaxy evolution. It is invoked to explain different properties of local massive galaxies, including their stellar mass function, in particular its higher mass end, the observed correlation between their stellar velocity dispersion and the mass of their supermassive black hole, and the enrichment of their circumgalactic medium (e.g., \citealt{silk98, fabian99, king03, springel05b, kormendy13, nelson19}). AGN feedback couples the energy that is released by the accreting black hole with the gas in the interstellar medium (ISM) of the host galaxy. 

The two main AGN feedback modes, which are related to the type of nuclear activity, are (i) the radiative/quasar mode, which is associated with black holes accreting close to the Eddington limit, and (ii) the kinetic/jet mode, which is associated with lower-power AGN with Eddington ratios of $\lesssim 0.1$ (e.g., \citealt{croton06, alexander12, fabian12, harrison18}). In the radiative feedback mode, the quasar is believed to drive gas outflows that can reach galactic scales, destroy, push, or even compress molecular clouds, and escape the host galaxy (e.g., \citealt{faucher12, zubovas12, zubovas14}). While some simulations suggest that these outflows can have a dramatic impact on their host, by violently quenching its star formation and transforming it into a red and dead elliptical (e.g., \citealt{dimatteo05, springel05, springel05b, hopkins06}), others suggest a more limited impact on the ISM, with these outflows escaping along the path of least resistance (e.g., \citealt{gabor14, hartwig18}). At the current stage, cosmological hydrodynamical simulations of galaxy formation can reproduce the observed properties of local massive galaxies only when AGN feedback is invoked, and only if AGN-driven winds carry a significant fraction of the AGN energy (1--10\% of $\mathrm{L_{AGN}}$; e.g., \citealt{crain15, pillepich18, dave19, nelson19, villaescusa_navarro21}).

Although AGN feedback is expected to be most prominent at the epoch of peak cosmic black hole accretion and star formation at z$\sim$2 (e.g., \citealt{merloni08, madau14}), observing AGN feedback in the local universe offers significant advantages in the resulting signal-to-noise ratio (SNR) and spatial resolution. Local AGN host galaxies are bright enough to allow the detection of weak outflow kinematic components in multiple gas phases (e.g., warm ionized, neutral atomic, warm molecular, and cold molecular; see review by \citealt{veilleux20}) in a relatively representative population of galaxies (e.g., \citealt{rupke05b, rupke05c, mullaney13, cicone14, woo16, fiore17, baron19b, mingozzi19, lutz20, shangguan20, riffel21, baron22, riffel23}). In addition, spatially-resolved spectroscopy in optical, near-infrared, and mm wavelengths can resolve these winds on scales ranging from tens to hundreds of parsecs in local galaxies (e.g., \citealt{liu13a, liu13b, cicone14, davies_r14, harrison14, husemann16, villar_martin16, rupke17, fluetsch19, mingozzi19, wylezalek20, fluetsch21, perna21, revalski21, ruschel_dutra21, deconto_machado22}). 

During the past decade, studies used large and public surveys (e.g., SDSS: \citealt{york00}; SAMI: \citealt{croom21}, and MANGA: \citealt{bundy15}), along with dedicated deeper multi-wavelength observations to study the occurrence rate, extent, kinematics, excitation/ionization, and density of multi-phased outflows in typical AGN hosts in the local universe. Using these properties, studies estimated the energy that is carried out by the multiphased winds, typically finding values that are several orders of magnitude lower than the theoretical coupling efficiency of 1--10\% (e.g., \citealt{villar_martin16, fiore17, rupke17, baron19b, fluetsch19, lutz20, shangguan20, fluetsch21, revalski21, ruschel_dutra21, baron22}). It is unclear whether the observations are in tension with the theoretical requirement, as simulations require a coupling efficiency of 1--10\% at the wind injection scale ($<<$kpc), while the observed winds are typically detected on kpc scales. According to one scenario, the wind evolves from injection scale to galactic scales, where it is shock-heated when encountering the host ISM, creating a bubble of extremely hot and ionized gas that dominates the energetics of the flow (e.g., \citealt{faucher12, richings18}). In such a case, observations that trace $T<10^{5}$ K gas will underestimate the total energy that is carried out by the wind. 

At higher redshifts of z$\sim$2, the mass and energetics of outflows in AGN hosts are still highly uncertain. Earlier studies focused on the most extreme objects or outflow cases, and were limited to small sample sizes and/or spatially-integrated observations. (e.g., \citealt{nesvadba08, cano12, perna15, brusa15}). Later works studied larger samples, including more typical high-redshift galaxies, using integral field unit (IFU) observations with adaptive optics in some cases (e.g., \citealt{newman12, genzel14, harrison16, forster_schreiber18, forster_schreiber19, leung19, kakkad20}; see review by \citealt{forster_schreiber20}). Despite these advances, the resulting sensitivity and spatial resolution do not allow to detect and resolve the weaker transitions of the outflow, making its reddening, ionization state, and density, still uncertain. As a result, the energy that is carried out by outflows at z$\sim$2 remains largely unconstrained. This is expected to change in the near future, with first observations by JWST/NIRSpec detecting and resolving z$\sim$2--4 outflows even in weak transitions such as [SII] and H$\beta$ emission lines and NaID absorption (e.g., \citealt{davies23, rupke23, veilleux23, wang24}). 

Another possible site for significant AGN feedback are galaxies on the merger sequence. During this short phase, the galaxies reach a peak in SFR and black hole accretion rate (e.g., \citealt{sanders88, sanders96}), leading to powerful supernova and AGN-driven winds (e.g, \citealt{springel05b, hopkins06}). A particular short phase within this sequence is the post-starburst phase (see review by \citealt{french21}). Post-starburst galaxies are believed to be galaxies in a rapid transition from starburst to quiescence, with optical spectra that are dominated by A-type stars, suggesting a significant burst of star formation that ended abruptly $<$1 Gyr ago \citep{dressler99, poggianti99, goto04, dressler04}. Some of these systems are bulge-dominated with tidal features, suggesting that they are merger remnants \citep{canalizo00, yang04, goto04, cales11}. Stellar population synthesis modeling of their optical spectra suggest high peak SFRs, ranging from 50 to 300 $\mathrm{M_{\odot}/yr}$ \citep{kaviraj07}, with estimated mass fractions forming in the burst of 10\% to 80\% of the total stellar mass \citep{liu96, norton01, yang04, kaviraj07, french18}.

\floattable
\begin{deluxetable}{c c c c c c c c c  c c c c}
\tablecaption{Galaxy properties and follow-up observations\label{tab:gal_properties}}
\tablecolumns{13}
\tablenum{1}
\tablewidth{0pt}
\tablehead{
\colhead{(1)} & \colhead{(2)} & \colhead{(3)} & \colhead{(4)} & \colhead{(5)} & \colhead{(6)} & \colhead{(7)} & \colhead{(8)} & \colhead{(9)} & \colhead{(10)} & \colhead{(11)} & \colhead{(12)} & \colhead{(13)} \\
\colhead{Object ID} &   \colhead{RA}  & \colhead{Dec} & \colhead{Plate} & \colhead{MJD} & \colhead{Fiber} & \colhead{z} & \colhead{$\log L_{\mathrm{bol}}$} & \colhead{$\log L_{\mathrm{SF}}(\mathrm{IR})$} & \colhead{Instrument} & \colhead{Mode}   & \colhead{resolution} & \colhead{$t_{\mathrm{exp}}$} \\
\colhead{} & \colhead{(deg)} & \colhead{(deg)} & \colhead{}  &  \colhead{} &  \colhead{} &  \colhead{} & \colhead{[erg/sec]} & \colhead{[$L_{\odot}$]} & \colhead{} & \colhead{}  & \colhead{(arc-sec)}  & \colhead{(hr)}  
}
\startdata
J022912   & 37.3041  & -5.1891 & 4388 & 55536 & 708 & 0.074 & 45.0 & 11.3$^{a}$  & MUSE       & WFM    & 0.8 & 2\\
J080427   & 121.1160 & 13.4918 & 2268 & 53682 & 278 & 0.134 & 44.7 & 11.0$^{a}$  & MUSE       & WFM-AO & 0.4 & 2\\
J112023   & 170.0996 & 15.7318 & 2494 & 54174 & 129 & 0.159 & 45.0 & 11.5$^{a}$  & MUSE       & WFM-AO & 0.4 & 2\\
J020022   & 30.0948  & 0.6566  & 403  & 51871 & 583 & 0.163 & 43.8 & 10.4$^{b}$  & MUSE       & WFM-AO & 0.4 & 2\\
J111943   & 169.9326 & 10.8436 & 1605 & 53062 & 122 & 0.177 & 44.8 & 11.4$^{a}$  & MUSE       & WFM-AO & 0.4 & 2\\
\hline
J003443   & 8.6820   & 25.1724 & 6281 & 56295 & 12  & 0.118 & 44.2 & 10.5$^{b}$  & KCWI      & medium+BM & 0.8 & 0.5\\
J124754   & 191.9789 & -3.6274 & 337  & 51997 & 285 & 0.090 & 45.0 & 11.3$^{b}$  & MUSE       & WFM    & 0.8 & 2\\
\enddata

\tablecomments{(1): Object ID throughout the paper. (2), (3): Right ascension and declination. (4), (5), (6): SDSS plate-mjd-fiber. (7): Redshift. (8) AGN bolometric luminosity. (9) star formation luminosity, estimated using: $^{a}$: a combination of far-infrared IRAS observations and ALMA or NOEMA mm observations, or $^{b}$: using a combination of IRAS observations and the estimated AGN luminosity. (10): IFU instrument. (11): Observing mode of IFU. (12): seeing/spatial resolution during observations. (13) On-source exposure time. The last two objects were studied in \citet{baron18} and \citet{baron20} respectively.}
\end{deluxetable}

Although AGN feedback is believed to be one of the processes responsible for the sudden quenching of star formation in post-starbursts, little is known observationally about AGN-driven winds in this short-lived phase. High velocity gas flows have been detected in post-starburst galaxies (e.g., \citealt{tremonti07, coil11, maltby19}), though in some cases were later attributed to winds driven by obscured starbursts in the systems (e.g., \citealt{diamond_stanic12}). To study the properties of AGN-driven winds in post-starburst galaxies, in \citet{baron22} we constructed a sample of local post-starburst candidates with AGN and evidence for ionized outflows. In \citet{baron18} and \citet{baron20} we used optical IFUs to spatially-resolve the multiphased gas in two such galaxies, finding massive ionized+neutral outflows with kinetic powers that are 10--100 times larger than those observed in typical active galaxies in the local universe. We therefore suggested that AGN feedback, in the form of galactic-scale outflows, may be significant in the post-starburst phase.

In this paper we present follow-up MUSE/VLT observations of 5 additional post-starburst galaxies with AGN and ionized outflows. Together with the two already-published post-starbursts (\citealt{baron18, baron20}), we perform a detailed analysis of their stellar population and multiphased gas, paying particular attention to the detection and characterization of galactic-scale flows. The paper is organized as follows. In section \ref{s:data} we describe our methods, including the sample selection (\ref{s:data:sample}), MUSE observations (\ref{s:data:observations}), stellar population analysis (\ref{s:data:stellar_pop}), ionized  (\ref{s:data:ionized_gas}) and neutral (\ref{s:data:neutral_gas}) gas analysis, collection of ancillary properties (\ref{s:data:ancillary}), and characterization of outflows (\ref{s:data:outflows}). In section \ref{s:results} we describe our results, and discuss their broader context in section \ref{s:discussion}. We conclude and summarize in section \ref{s:conclusions}. Readers who are interested only in the results may skip directly to section \ref{s:results}. Throughout this paper we use a Chabrier initial mass function (IMF; \citealt{chabrier03}), and assume a standard $\Lambda$CDM cosmology with $\Omega_{\mathrm{M}}=0.3$, $\Omega_{\Lambda}=0.7$, and $h=0.7$.

\section{Methods}\label{s:data}

\begin{figure*}
	\centering
\includegraphics[width=0.78\textwidth]{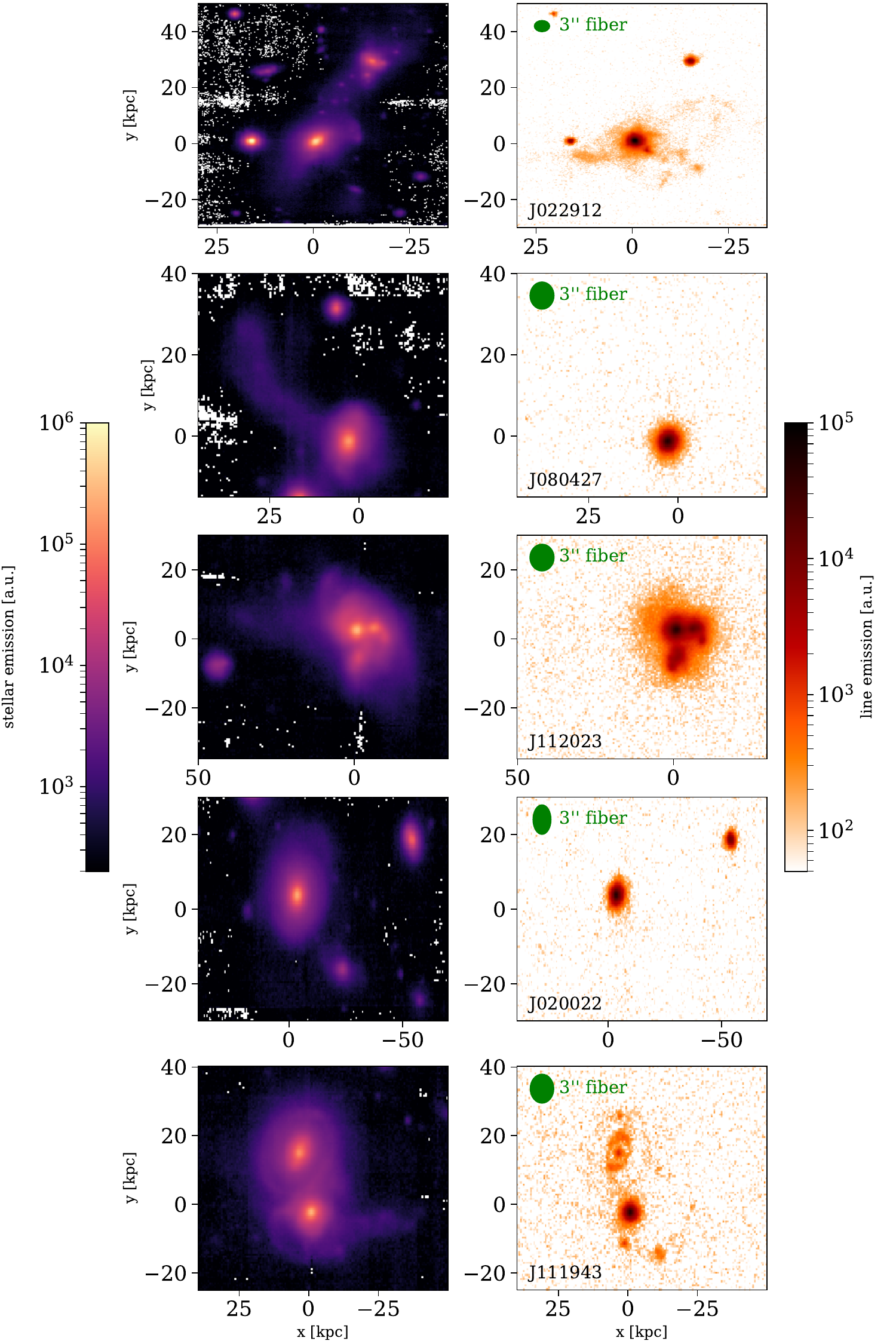}
\caption{\textbf{The MUSE view of the stellar and gas emission of the galaxies in our sample.} Each row represents a galaxy from our sample. The left column shows the integrated emission within the rest-frame wavelength range 5300--5700 \AA, assuming the redshift of the target galaxy. The right column shows the integrated continuum-subtracted flux in the rest-frame wavelength range 6520--6620 \AA, which includes the H$\alpha$ and [NII] lines. The 3'' SDSS fiber is marked as a green circle. 
}\label{f:stellar_and_gas_emission}
\end{figure*}

\subsection{Sample selection}\label{s:data:sample}

Our sample was selected from our parent sample of post-starburst galaxy candidates with AGN and ionized outflows, and is described in detail in \citet{baron22}. For completion, we give a brief overview of its main properties. The parent sample was drawn from the 14th data release of the Sloan Digital Sky Survey (SDSS; \citealt{york00}). To select post-starburst systems, we selected galaxies with strong H$\delta$ absorption lines, in particular requiring $\mathrm{EW(H\delta) > 5}$\AA, where EW is the absorption equivalent width. To select post-starbursts with AGN, we performed emission line decomposition on the stellar continuum-subtracted spectra, fitting narrow and broad kinematic components to the emission lines: H$\alpha$, H$\beta$, [OIII], [NII], [SII], and [OI]. We selected systems with narrow line ratios that are consistent with AGN photoionization using standard line-diagnostic diagrams (including LINERs; \citealt{kewley01, cidfernandes10}). We used the broad Balmer emission lines to filter out type I AGN. To select post-starbursts with AGN and ionized outflows, we selected systems in which broad kinematic components are detected in both the Balmer and forbidden lines. We found a total of 215 post-starburst candidates with AGN and ionized outflows, out of which 144 show evidence for an ionized outflow in multiple lines.

We selected a subset of 32 systems for follow-up observations. These galaxies show the highest SNRs in the broad kinematic components of the [OIII] and H$\alpha$ emission lines. Such a selection is biased towards galaxies with more luminous emission lines, and thus favors systems with higher AGN luminosity and SFR. In addition, as we show throughout the paper, it may also favor tidally-interacting systems with significant interaction-induced flows. We followed-up a subset of 15 galaxies with NOEMA, and report the result in \citet{baron23}. One of the galaxies was observed with KCWI/Keck and results are reported in \citet{baron18}. Another galaxy was observed with MUSE/VLT and results are reported in \citet{baron20}. In this work we present MUSE/VLT observations of 5 additional galaxies from this subset, 4 of which are part of the NOEMA sample from \citet{baron23}.

\subsection{MUSE observations}\label{s:data:observations}

MUSE is a second generation integral field spectrograph on the VLT \citep{bacon10}. It consists of 24 integral field units that cover the wavelength range 4650\AA--9300\AA, achieving a spectral resolution of 1750 at 4650\AA\, to 3750 at 9300\AA. In its Wide Field Mode (WFM), MUSE splits the 1$\times$1 arcmin$^{2}$ field of view (FOV) into 24 channels which are further sliced into 48 15''$\times$0.2'' slitlets. Since period 101, MUSE uses the GALACSI adaptive optics (AO) module to offer AO corrected WFM, WFM-AO. 

The five galaxies were observed as part of our program "Mapping AGN-driven outflows in quiescent post starburst E+A galaxies" (0100.B-0517(A) and 0102.B-0228(A)). For the earlier program, 0100.B-0517(A), observations were carried out in a seeing-limited WFM with a pixel scale of 0.2'' and a spatial resolution of $\sim$0.8''. For the later program 0102.B-0228(A), observations were carried out with WFM-AO, reaching spatial resolution of 0.4''. We downloaded the data from the ESO phase 3 online interface, which provides fully reduced, calibrated, and combined MUSE data for all targets with multiple observing blocks. Table \ref{tab:gal_properties} lists the galaxies' properties and follow-up observations, including the instruments, modes, and exposure times for each source. In figure \ref{f:stellar_and_gas_emission} we show the stellar continuum and ionized gas emission for the five galaxies in our sample.

\begin{figure*}
	\centering
\includegraphics[width=1\textwidth]{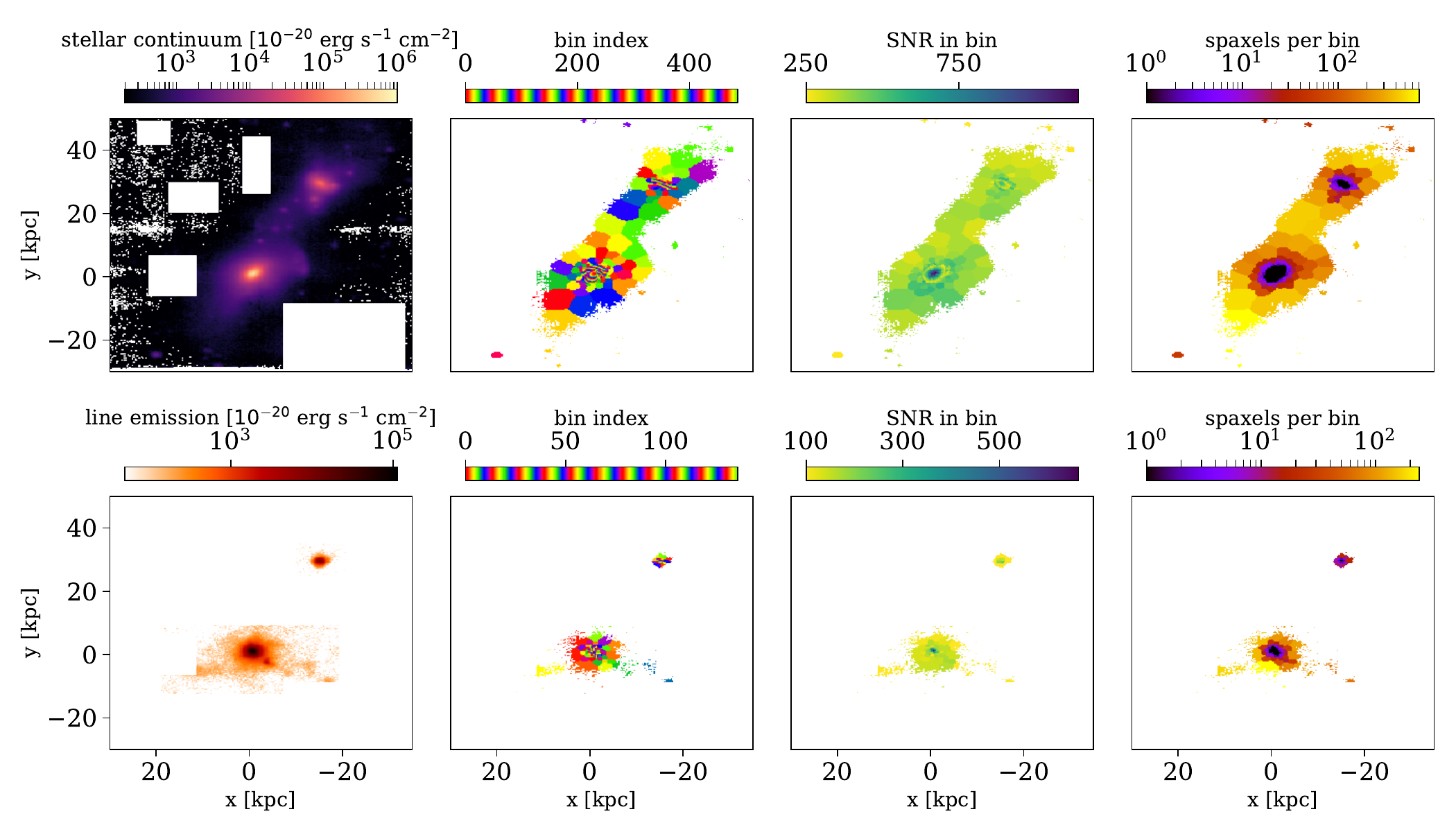}
\caption{\textbf{Illustration of the Voronoi binning applied to the data cubes.} The top row represents the binning done for the stellar properties (section \ref{s:data:stellar_pop}), and the bottom for the emission and absorption lines (sections \ref{s:data:ionized_gas} and \ref{s:data:neutral_gas}). The first column shows the integrated flux, which is used as the signal $S$ when defining the spaxel SNR. For the stellar properties, the wavelength range is 5300--5700 \AA, and for the gas properties, the summed flux is the continuum-subtracted 6520--6620 \AA\, range, which includes the H$\alpha$ and [NII] lines. The second column shows the bin indices assigned to each spaxel, and the third shows the combined SNR of each bin. By construction, the minimal combined SNR is 250 for the stellar binning and 100 for the line binning, since these are the target SNRs we selected. The forth column shows the number of spaxels in every bin. The Voronoi bins include only single spaxels in the centers of the galaxies, while having tens of spaxels in the outskirts. 
}\label{f:voronoi_binning_example_for_J022912}
\end{figure*}

\subsection{Stellar properties}\label{s:data:stellar_pop}

To accurately determine the stellar kinematics and stellar population properties from full spectral fitting, a minimum SNR is required (\citealt{johansson12, westfall19}). We therefore binned the spaxels using {\sc vorbin}, which is a broadly-used package to bin data cubes along the two spatial dimensions using Voronoi tessellations \citep{cappellari03}. The method uses an iterative process to find the optimal binning solution, given the SNR of each individual spaxel and the target SNR as inputs. We concentrated on the wavelength range 5300--5700 \AA, which is free from strong emission or absorption lines. We defined the SNR of an individual spaxel to be $S/\sigma$, where $S$ is the sum of the flux in this wavelength range, and $\sigma$ is the square-root of the sum of squared residuals of a 2-degree polynomial fit in this range. The fit to a 2-degree polynomial is required to account for large-scale continuum variations between 5300 and 5700 \AA. 

Prior to applying {\sc vorbin} to the data cubes, we manually inspected the cubes and identified all the objects that are not associated with the primary galaxy (i.e., high-z galaxies or stars in the field), and masked them out. Galaxies that show a similar redshift to the primary galaxy, or large-scale tidal tails, were not masked out. We then manually inspected the spectra of spaxels with different SNRs of the collapsed cubes and found that spaxels with SNR $<10$ generally do not show any signal that can be associated with the galaxy, and instead dominated by sky lines. We therefore masked these spaxels out. As shown in figure \ref{f:voronoi_binning_example_for_J022912}, this step only masks out spaxels that are far away ($\gtrapprox$30 kpc) from the galaxy (or pair). We then applied {\sc vorbin} to the masked data cubes, experimenting with different target SNRs. We found that requiring a target SNR of 250 of the collapsed cubes allows us to keep high spatial resolution in the centers of the galaxies, while also allowing us to trace the stellar properties accurately (best fitting stellar parameters are estimated with $>3\sigma$) in their outskirts. The resulting stellar properties do not change significantly for target SNRs between 150 and 350. Table \ref{tab:voronoi_bins_parameters} in \ref{a:voronoi_bins} summarizes the binning parameters and number of bins per galaxy. The top row of figure \ref{f:voronoi_binning_example_for_J022912} illustrates the binning process for J022912. The figure shows that the Voronoi bins include only single spaxels in the center of the galaxy, while including tens of spaxels in the outskirts. 

To estimate the stellar kinematics and population properties, we used {\sc ppxf} \citep{cappellari12} to fit stellar population synthesis models to the binned spectra. {\sc ppxf} is a widely-used code for extraction of stellar kinematics and stellar population from absorption line spectra of galaxies \citep{cappellari04}. We used the MILES library, which contains single stellar population synthesis models that cover the entire optical wavelength range with a FWHM resolution of 2.3 \AA\, \citep{vazdekis10}. We used models produced with the Padova 2000 stellar isochrones assuming a Chabrier IMF \citep{chabrier03}. The stellar ages of the MILES models range between 0.03 to 14 Gyr. Since we are also interested in the dust reddening towards the stars, we applied {\sc ppxf} without any additive or multiplicative Legendre polynomials. The output of {\sc ppxf} includes the best-fitting stellar model and $\chi^2$, the contribution of stars with different ages (i.e., a non-parameteric star formation history), the stellar kinematics, and the reddening towards the stars, assuming a \citet{calzetti00} extinction law.

\subsection{Ionized gas}\label{s:data:ionized_gas}

To study the properties of the ionized gas, we first need to fit and subtract the stellar continuum emission. The Voronoi bins obtained in section \ref{s:data:stellar_pop} for the stellar population are optimized for stellar continuum emission rather than for line emission. For example, figure \ref{f:stellar_and_gas_emission} shows significant line emission in regions where no stellar continuum emission is detected for J022912 and J111943. Such spaxels are masked out in section \ref{s:data:stellar_pop} prior to applying the Voronoi binning and thus will be missed. To achieve optimal spatial resolution and line SNR, we ran {\sc vorbin} again, using a different definition for spaxel SNR as described below. We concentrated on the wavelength range 6500--6640 \AA, which includes the H$\alpha$ and [NII] lines. We defined the `line region' to be 6520--6620 \AA\, and the `continuum region' to be 6500--6520 \AA\, and 6620--6640 \AA. We fitted the continuum region with a 1-degree polynomial, and subtracted it from the line region spectrum. The signal, $S$, is then the sum over the continuum-subtracted line region, and the noise, $\sigma$, is the square-root of the sum of squared residuals, multiplied by a factor $N_{\mathrm{line}}/N_{\mathrm{cont}}$, where $N_{\mathrm{line}}$ is the length of the line region and $N_{\mathrm{cont}}$ is the length of the continuum region. Therefore, this SNR definition is proportional to the integrated H$\alpha$+[NII] line flux. 

Prior to applying {\sc vorbin}, we masked out spaxels that were not spatially-associated with the galaxy (or galaxy pair; see all the scattered orange pixels throughout the FOV in figure \ref{f:stellar_and_gas_emission}). We experimented with different minimum SNRs of the integrated flux, below which spaxels were masked out, and target SNRs. The selected minimum and target SNRs are different for the different galaxies and are listed in table \ref{tab:voronoi_bins_parameters} in \ref{a:voronoi_bins}. The minimum SNRs are not significantly different for different galaxies, and range between 5 and 8. Changing the minimum SNR mostly affects the number of spaxels included within bins in the outskirts of the galaxies, where tens of spaxels are binned together. As for the target SNR, there is a tradeoff between setting a lower and a higher value. For a higher value, the resulting line SNR is higher, allowing the detection of broad kinematic components of weak lines such as H$\beta$ and [SII]. However, a higher value also results in larger bins, often mixing separate gas kinematic components into a single bin. The latter, for example, results in double or triple-peaked line profiles which are more challenging to interpret. Our selected target SNRs, which range between 30 and 100, were chosen as the maximal SNRs for which different kinematic components remain not blended. Therefore, most of the binned spectra exhibit either one (narrow) or two (narrow + broad) kinematic components\footnote{The exception is J022912, for which several central spaxels include two narrow components (double-peaked profile) and a broad kinematic component. However, the bins in the center of the galaxy include single spaxels.}. The bottom row of figure \ref{f:voronoi_binning_example_for_J022912} illustrates the binning process for J022912, and shows that the bins include single spaxels in the galaxy center and tens of spaxels in the outskirts. 

We then used {\sc ppxf} to fit and subtract the stellar continuum from the binned spectra. Our emission line decomposition follows the method outlined in \citet{baron20}, which is briefly summarized below. We fit the emission lines H$\beta$, [OIII]\footnote{For J112023, the 5007 \AA\, [OIII] component is blocked by the reduction pipeline since it coincides with the WFM-AO laser's sodium lines. The 4959 \AA\, [OIII] is not masked out and we fit it. For J020022 and J111943, both of the [OIII] lines are masked out and we do not fit them.}, [OI], [NII], H$\alpha$, and [SII] in each binned spectrum. The amplitude ratios of the [NII] and [OIII] doublet lines are set to their theoretical values. The [SII]$\lambda$6717\AA/[SII]$\lambda$6731\AA\, intensity ratio is allowed to vary between 0.44 and 1.44. Similarly to the post-starburst studied in \citet{baron20}, the galaxies in our sample are dusty and thus their H$\alpha$+[NII] lines are significantly stronger than their H$\beta$+[OIII]. We therefore started by fitting the H$\alpha$ and [NII] lines.  We modeled each of the emission lines using one or two Gaussians, where the first represents the narrow kinematic component and the second the broader kinematic component\footnote{An exception is J022912, for which the central spaxels show a double-peaked narrow kinematic component and a broad one. For this source, we also examined a model with two narrow Gaussians and one broad, and used its best-fitting parameters if its reduced $\chi^{2}$ was lower than the $\chi^{2}$ obtained for the one narrow and one broad model.}. We tied the central wavelengths and widths of the narrow Gaussians to have the same velocity and velocity dispersion, and did the same for the broader Gaussians. The broad kinematic component was kept if the reduced $\chi^{2}$ was improved compared to a fit with only a narrow kinematic component, and only if its flux in H$\alpha$ and [NII]6584\AA\, was detected to more than 3$\sigma$. 

Once we obtained a fit for the H$\alpha$+[NII] complex, we used the best-fitting kinematic parameters to fit the other ionized emission lines. We fitted H$\beta$, [OIII], [OI], and [SII] with narrow and broad (if exists in the H$\alpha$+[NII] fit) kinematic components, locking their central wavelengths and line widths to the best-fitting values obtained for the H$\alpha$+[NII]. A broad kinematic component was considered detected only if the amplitude of the broad Gaussian was larger than 3 times its uncertainty. Otherwise, we ran the fit again with a narrow kinematic component only. 

To estimate the line fluxes, we integrated over the best-fitting profiles. We assume that the narrow kinematic component originates from non-outflowing gas in these galaxies. We therefore integrated over the best-fitting narrow profiles to obtain the line fluxes associated with the non-outflowing gas. As discussed in section \ref{s:results:mergers}, most of the galaxies in our sample exhibit signs of ongoing mergers and/or disturbed gas kinematics. Broad kinematic components are detected in a large fraction of the spaxels, and it is not clear whether these components originate from outflowing gas or from gas disturbed by the merger. We therefore followed the conservative approach by \citet{lutz20}, and estimated the flux in the wings of the broad component, where we integrated the broad component only over wavelengths in which it contributes more than 50\% of the total flux density of the line (i.e., wavelengths in which the flux density of the broad component is larger than the flux density of the narrow component). The broad component dominates the line profile for velocities in the range $\pm$(600--1200) km/sec with respect to the narrow core. 

We then used the measured H$\alpha$/H$\beta$ flux ratios to estimate the dust reddening towards the line-emitting gas, once for the non-outflowing gas and once for the broad wings. Assuming case-B recombination, a gas temperature of $10^{4}$ K, a dusty screen, and the \citet{cardelli89} extinction law, the color excess is given by:
\begin{equation}\label{eq:reddening}
	{\mathrm{E}(B-V) = \mathrm{2.33 \times log\, \Bigg[ \frac{(H\alpha/H\beta)_{obs}}{2.85} \Bigg] \, \mathrm{mag} }},
\end{equation}
where $\mathrm{(H\alpha/H\beta)_{obs}}$ is the observed line ratio. We then corrected all the observed fluxes for dust extinction using the derived $\mathrm{E}(B-V)$ values. 

We used the [SII]$\lambda$6717\AA/[SII]$\lambda$6731\AA\, intensity ratio to estimate the electron density in the gas, once for the narrow component and once for the broad wings. Assuming gas temperature of $10^{4}$ K, the electron density is given by (e.g., \citealt{fluetsch21}):
\begin{equation}\label{eq:elec_density}
	{n_{\mathrm{e}} = \frac{cR - ab}{a - R}},
\end{equation}
where $R$ is [SII]$\lambda$6717\AA/[SII]$\lambda$6731\AA, and $a=0.4315$, $b=2107$, $c=627.1$. Due to the critical densities of the two [SII] transitions, the intensity ratio is sensitive to electron densities in the range 50--5000 $\mathrm{cm^{-3}}$, and its value ranges between 0.44 and 1.44. For electron densities outside this range, the intensity ratio is constant and cannot be used to infer the density. In addition, in \citet{baron19b} and \citet{davies20} we used the ionization parameter of the gas to estimate the electron density, and found that the [SII] lines can underestimate the electron density in the ionized outflow. By comparing the observed line ratios to photoionization models, we suggested that this is because the [SII] lines are emitted close to the ionization front of the cloud rather than the mean over the ionized region. Since the electron density drops rapidly near the ionization front, the [SII] lines trace low electron density regions. Therefore, we also use the ionization parameter method presented in \citet{baron19b} to estimate the electron density.
\subsection{Neutral gas}\label{s:data:neutral_gas}

To study the neutral gas properties, we used the same Voronoi bins defined in section \ref{s:data:ionized_gas} for the emission lines. The binned spectra show evidence for NaID absorption, emission, or a combination of the two. Interstellar NaID absorption, in particular blueshifted absorption, has been detected in numerous starburst and AGN-dominated systems. It has been widely used to constrain the neutral phase of galactic winds (see a review by \citealt{veilleux20}). Redshifted NaID emission, until recently, has only been detected in a handful of sources (\citealt{rupke15, perna19, baron20}). However, more recent studies using 1D spectroscopy \citep{baron22} and IFU observations \citep{fluetsch21} of large samples of starburst and AGN-dominated galaxies detect NaID emission in a large fraction of the sources and/or spaxels ($\gtrsim$30\%), suggesting that NaID emission is not a rare phenomenon.

NaID absorption is the result of absorption of continuum photons along the line of sight. On the other hand, NaID emission is the result of absorption by neutral Sodium outside the line of sight and an isotropic reemission. As we discussed extensively in \citet{baron20} and \citet{baron22}, a neutral outflow may produce a P-Cygni profile in the NaID lines, with the approaching part of the outflow producing blueshifted absorption, and the receding part of the outflow producing redshifted emission (see also \citealt{prochaska11}). Neglecting the redshifted NaID emission component may result in an underestimation of the neutral outflowing gas mass, similarly to neglecting the red wing of broad emission lines originating in ionized outflows.

Our modeling of the NaID absorption and emission profile follows the methodology outlined by \citet{baron22}. We refer the reader to appendix B in that paper for the full model, discussion of assumptions, and possible degeneracies between the parameters, while here we only briefly summarize the main components of the model. We considered three different models: (i) NaID absorption only, (ii) NaID emission only, and (iii) a combination of blueshifted NaID absorption and redshifted emission. To properly model the NaID profile, we must also include a model for the HeI$\lambda$5876\AA\, emission. We modeled the HeI profile using the best-fitting parameters of the H$\alpha$ line, where the central wavelength and line width were locked to those of the H$\alpha$, and the HeI/H$\alpha$ amplitude ratio was allowed to vary between 0.01 and 0.05. In the most general case of NaID absorption and emission, we modeled the observed flux as:
\begin{equation}\label{eq:naid_model}
	\begin{split}
	& f(\lambda) = \Big[f_{\mathrm{stars}}(\lambda) + f_{\mathrm{HeI}}(\lambda) + f_{\mathrm{NaID\, emis}}(\lambda) \Big] \times I_{\mathrm{NaID\, abs}}(\lambda),
	\end{split}
\end{equation}
where $f(\lambda)$ is the observed flux, $f_{\mathrm{stars}}(\lambda)$ is the stellar continuum obtained using {\sc ppxf}, $f_{\mathrm{HeI}}(\lambda)$ represents the HeI emission, $f_{\mathrm{NaID\, emis}}(\lambda)$ represents the redshifted NaID emission, and $I_{\mathrm{NaID\, abs}}(\lambda)$ represents the NaID absorption. Since the NaID emission is additive, while the NaID absorption is multiplicative, one must model the observed spectrum rather than the normalized one. We fitted all three models and selected the model with the lowest reduced $\chi^{2}$. In figure \ref{f:NaID_fitting_example} in \ref{a:NaID_fitting} we show examples of the best-fitting NaID profile in cases of absorption only, emission only, and a combination of the two. 

In case of NaID absorption, we used the best-fitting optical depth $\mathrm{\tau_{0}(NaID_{K})}$ to estimate the neutral Sodium column density via (e.g., \citealt{draine11}):
\begin{equation}\label{eq:NaID_column}
	\begin{split}
	& N \mathrm{_{NaI} = 10^{13}\, cm^{-2}} \times \\
	& \mathrm{ \Big( \frac{\tau_{0}(NaID_{K})}{0.7580} \Big) \Big(\frac{0.4164}{f_{lu}} \Big) \Big(\frac{1215\,\AA}{\lambda_{lu}}\Big) \Big(\frac{b}{10\,km/sec}\Big)},
	\end{split}
\end{equation}
where $\mathrm{f_{lu}}=0.32$, $\mathrm{\lambda_{lu}=5897}$ \AA, and $b$ is the Doppler parameter, which is related to the velocity dispersion via $b = \sqrt{2} \sigma$. We then estimated the Hydrogen column density using (\citealt{shih10}):
\begin{equation}\label{eq:Hydrogen_column}
	{N_{\mathrm{H}} = \frac {N_{\mathrm{NaI}}} {(1 - y) 10^{A + B+ C}}  },
\end{equation}
where $(1 - y)$ is the Sodium neutral fraction which we assume to be 0.1, $A$ is the Sodium abundance term, $B$ is the Sodium depletion term, and $C$ is the gas metallicity term. Following \citet{shih10}, we took $A = \log [N_{\mathrm{Na}}/N_{\mathrm{H}}] = -5.69$ and $B = \log [N_{\mathrm{Na}} / N_{\mathrm{H, total}}] - \log [N_{\mathrm{Na}} / N_{\mathrm{H, gas}}] = -0.95$. For the stellar masses of the systems in our sample, the mass-metallicity relation (e.g., \citealt{t04}) suggests that the metallicity is roughly twice solar. We therefore used $C = \log [Z/ 1\, Z_{\odot}] = \log [2]$.

In case of redshifted NaID emission, we integrated the emission line profile to obtain the flux. The NaID-emitting gas has a comparable spatial extent to the broad wings-emitting gas (see figure \ref{f:J022912_NaID_emis_and_abs_rep} and section \ref{s:results:neutral}). We therefore assume that it is affected by roughly similar dust columns, and correct the NaID flux for dust extinction using the $\mathrm{E}(B-V)$ derived using the H$\alpha$/H$\beta$ flux ratio in the broad wings.

\subsection{Ancillary properties}\label{s:data:ancillary}

We extracted the stellar masses reported in the MPA-JHU value added catalogue for the galaxies in our sample \citep{kauff03b, t04}. For the AGN bolometric luminosity, we estimated the total dust-corrected H$\beta$ luminosity by integrating the luminosities of all the spaxels that are associated with the primary galaxies and with line ratios consistent with AGN ionization (see figure \ref{f:BPT_narrow_and_wings}). We then used the bolometric correction factor from \citet{netzer19} to convert the dust-corrected H$\beta$ luminosity to bolometric luminosity.

We use several different estimates for the star formation rate (SFR). In \citet{baron22} we used IRAS 60 $\mathrm{\mu m}$ observations to show that many systems selected to have post-starburst optical signatures host in fact significant obscured star formation. In particular, for our parent sample of post-starbursts with AGN and ionized outflows, we found that 45\% are $>0.3$ dex above the star-forming main sequence (confidence intervals 36\%--56\%), and 32\% are $>0.6$ dex above (confidence intervals 24\%--41\%). We found a significant correlation between the far-infrared SFR and the AGN bolometric luminosity, which is in line with the relation observed in active starbursts. In \citet{baron23} we used NOEMA observations to study the star formation and molecular gas properties in a subset of the galaxies. In particular, we combined the mm continuum emission from NOEMA with the IRAS 60 $\mathrm{\mu m}$ observations to estimate the SFR. Four out of the five galaxies were observed with NOEMA and we use the SFR estimates (or upper limits) reported by \citet{baron23}. The fifth galaxy has only an IRAS-based upper limit on the SFR. In total, 3/5 of the galaxies have SFR estimates, while 2/5 have only upper limits. We also used the derived AGN bolometric luminosities and our best fitting relation between L(AGN) and L(SF) from \citet[figure 5]{baron22} to estimate the far-infrared SFR. 

We list the AGN luminosities the SFRs in table \ref{tab:gal_properties}.

\subsection{Outflow properties}\label{s:data:outflows}

In this section we describe our methods to derive different outflow properties, starting with the ionized outflow. Assuming that the broad wings of the emission lines originate from an ionized outflow (see, however, section \ref{s:results:ionized}), we used the best-fitting line profiles to derive the outflow extent, velocity, ionization parameter, electron density, outflowing gas mass, mass outflow rate, and outflow kinetic power. 

As described in section \ref{s:data:ionized_gas}, broad kinematic components are detected in a large fraction of the spaxels in each galaxy. Since the galaxies are undergoing mergers and/or show signatures of disturbed gas kinematics, we followed the conservative approach by \citet[see figure 1 there]{lutz20} and considered only the wings of the broad profiles as originating from an outflow. The wings are defined as the wavelengths in which the broad component contributes more than 50\% of the total flux density. In each spaxel, we considered a red/blue wing detected if its integrated flux was larger than 3 times its uncertainty, which was estimated by propagating the uncertainties of the best-fitting parameters of the broad kinematic component. As a result, in some spaxels we detected only the red or only the blue wing. 

For the outflow extent, we defined the brightest spaxel in the primary galaxy as the center, and estimated the distance of all the spaxels in which broad wings in H$\alpha$ were detected from the center. We considered two definitions for the outflow extent. The first is $r_{\mathrm{outflow}} = r_{95}$, where $r_{95}$ is the 95th percentile of the distance distribution. According to this definition, the outflow extent is close to the maximal distance in which broad wings are detected. The second definition is the H$\alpha$ flux-weighted average distance, which is smaller than the maximal distance by a factor of 1--2.5. As discussed in section \ref{s:results:outflows}, our main conclusions do not change when adopting one definition versus the other. 

For the outflow velocity, we defined the maximal outflow velocity in each spaxel as $\Delta v + 2\sigma$ for the red wing and $\Delta v - 2\sigma$ for the blue wing, where $\Delta v$ and $\sigma$ are the centroid velocity and velocity dispersion of the broad kinematic component. We used these velocities when estimating the mass outflow rate and kinetic power in each spaxel. We also estimated the global outflow velocity in each galaxy as the H$\alpha$ flux-weighted average of the outflow velocities of the individual spaxels. We show in section \ref{s:results:ionized} that the red and blue wings are roughly symmetric in terms of dust-corrected H$\alpha$ flux, maximal velocity, and spatial extent. Therefore, for spaxels where both the red and blue wings are detected, the outflow velocity was defined as the average of the two. Otherwise, we adopted the velocity of the detected wing only.

We used the ionization parameter method presented in \citet{baron19b} to estimate the average electron density in the ionized wind independently from the [SII] method (equation 4 there). The ionization parameter method assumes AGN-photoionized gas and is based on the simple relation between the AGN luminosity, the gas distance from the AGN, and its ionization state. It requires knowledge of the AGN bolometric luminosity, the outflow extent, and its ionization parameter. For each galaxy, we used the median emission line ratios [NII]/H$\alpha$ and [OIII]/H$\beta$ in the broad wings to estimate the ionization parameter in the outflowing gas (equation 2 in \citealt{baron19b}). Using the estimated AGN bolometric luminosity (section \ref{s:data:ancillary}) and the outflow extent $r_{\mathrm{outflow}} = r_{95}$, we estimated the electron density in the outflow. 

We estimated the outflowing ionized gas mass $M_{\mathrm{ion}}$, mass outflow rate $\dot{M}_{\mathrm{ion}}$, and kinetic power $\dot{E}_{\mathrm{ion}}$ using equation 7 and the related text from \citet{baron19b}. These estimates are standard and have been used extensively in the literature (e.g., \citealt{harrison14, fiore17, rupke17, fluetsch21, ruschel_dutra21}). They require the knowledge of the dust-corrected H$\alpha$ luminosity, the outflow extent, the electron density, and the effective outflow velocity. We estimated these properties for the red and blue wings in each spaxel separately, and then obtained the global $M_{\mathrm{ion}}$, $\dot{M}_{\mathrm{ion}}$, and $\dot{E}_{\mathrm{ion}}$ by summing over all spaxels. 

\floattable
\begin{deluxetable}{c l l l l}
\tablecaption{Stellar and gas morphologies\label{tab:morphologies}}
\tablecolumns{5}
\tablenum{2}
\tablewidth{0pt}
\tablehead{
\colhead{Object ID} & \colhead{Morphology}                                                  & \colhead{Stellar kinematics} & \colhead{Narrow gas}     & \colhead{NaID profile}}
\startdata
J022912   & Companion at d$\sim$30 kpc and a bridge/tidal feature between.  & disk-like          & disturbed       & absorption \& emission   \\
J080427   & No visible companion. Tidal tail extending to $\sim$30 kpc.     & disk-like          & disturbed      & absorption               \\
J112023   & Ongoing merger with two nuclei and tidal features.              & disturbed          & disturbed      & absorption \& emission   \\
J020022   & Companion at d$\sim$50 kpc.                                     & disk-like          & disk-like       & no                       \\
J111943   & Companion at d$\sim$10 kpc and tidal features.                  & disk-like          & disk-like       & no                       \\
\hline
J003443   & Companion at d$\sim$3 kpc.                                      & -                  & -               & -                        \\
J124754   & Ordered spiral structure.                                       & disk-like          & disturbed      & absorption \& emission    \\
\enddata
\end{deluxetable}

For the neutral outflow, the best-fitting parameters of the NaID profile (see section \ref{s:data:neutral_gas}) include the absorption optical depth $\mathrm{\tau_{0}(NaID_{K})}$, covering factor $C_{f}$, centroid velocity, and velocity dispersion. If redshifted NaID emission is detected, then the best-fitting parameters also include the amplitude of the $\mathrm{NaID_{K}}$ emission component, the doublet amplitude ratio, centroid velocity, and velocity dispersion. To distinguish between absorption/emission that originate from a neutral outflow versus from the non-outflowing interstellar medium, we used the best-fitting centroid velocities. We considered the absorption to originate from an outflow if its centroid velocity is blueshifted by more than 100 km/sec from the centroid velocity of the stars. Similarly, NaID emission was considered to originate from an outflow if its centroid velocity is redshifted by more than 100 km/sec from the stars (see figures \ref{f:J022912_NaID_abs_properties_edited} and \ref{f:J022912_NaID_emis_properties_edited} in section \ref{s:results:neutral}).

We defined the maximal velocity of the neutral outflow to be $\Delta v - 2\sigma$ for the blueshifted absorption and $\Delta v + 2\sigma$ for the redshifted emission. Similarly to the ionized outflows, for each spaxel where an outflow is detected, we estimated its distance from the brightest spaxel. For the blueshifted absorption, to estimate the outflowing neutral gas mass $M_{\mathrm{neut}}$, mass outflow rate $\dot{M}_{\mathrm{neut}}$, and kinetic power $\dot{E}_{\mathrm{neut}}$, we assumed the thin shell model by \citet{rupke05b}, and used equations 6 and 7 from \citet{shih10}. These equations are the standard method to estimate the mass and energetics of the neutral outflow using spatially-resolved observations (see review by \citealt{veilleux20}).

To the best of our knowledge, the redshifted NaID emission has not been included in estimates of the mass and energetics of neutral outflows so far. To estimate the neutral gas mass that is associated with the NaID emission, we estimated the dust-corrected NaID luminosity in each spaxel where a redshifted outflow has been identified. The NaID emission line luminosity can be expressed as: $L_{\mathrm{emis}}(\mathrm{NaID}) = \mathrm{EW_{emis}(NaID)} \times L_{\lambda, \mathrm{stars}}$, where $L_{\lambda, \mathrm{stars}}$ is the stellar continuum at the NaID wavelength. We used the observed NaID luminosity and stellar continuum to estimate $\mathrm{EW_{emis}(NaID)}$. Assuming absorption on the linear part of the curve of growth (optically-thin absorption henceforth), the NaID column density associated with the redshifted emission is given by (\citealt{draine11}):
\begin{equation}\label{eq:NaID_column_emission}
	{N_{\mathrm{NaI, emis}} = 1.130 \times 10^{12}\, \mathrm{cm^{-2}} \frac{\mathrm{EW_{emis}(NaID)}}{\mathrm{f_{lu}} \lambda_{\mathrm{lu}}^2}},
\end{equation}
where $\mathrm{f_{lu}}=0.32$ and $\mathrm{\lambda_{lu}=5897}$. In section \ref{s:results:neutral} (see figure \ref{f:J022912_NaID_emis_properties_edited}) we show that for the NaID emission that is associated with the outflow, the NaID amplitude ratio is close to 2, suggesting optically-thin gas. We then used equation \ref{eq:Hydrogen_column} to convert the NaI column density to Hydrogen column density $N_{\mathrm{H, emis}}$. We consider $N_{\mathrm{H, emis}}$ as the Hydrogen column of the NaID-emitting gas, and use equations 6 and 7 from \citet{shih10} to estimate the gas mass, mass outflow rate, and kinetic power, of the neutral gas that produces the redshifted NaID emission.

\section{Results}\label{s:results}

\begin{figure*}
	\centering
\includegraphics[width=0.85\textwidth]{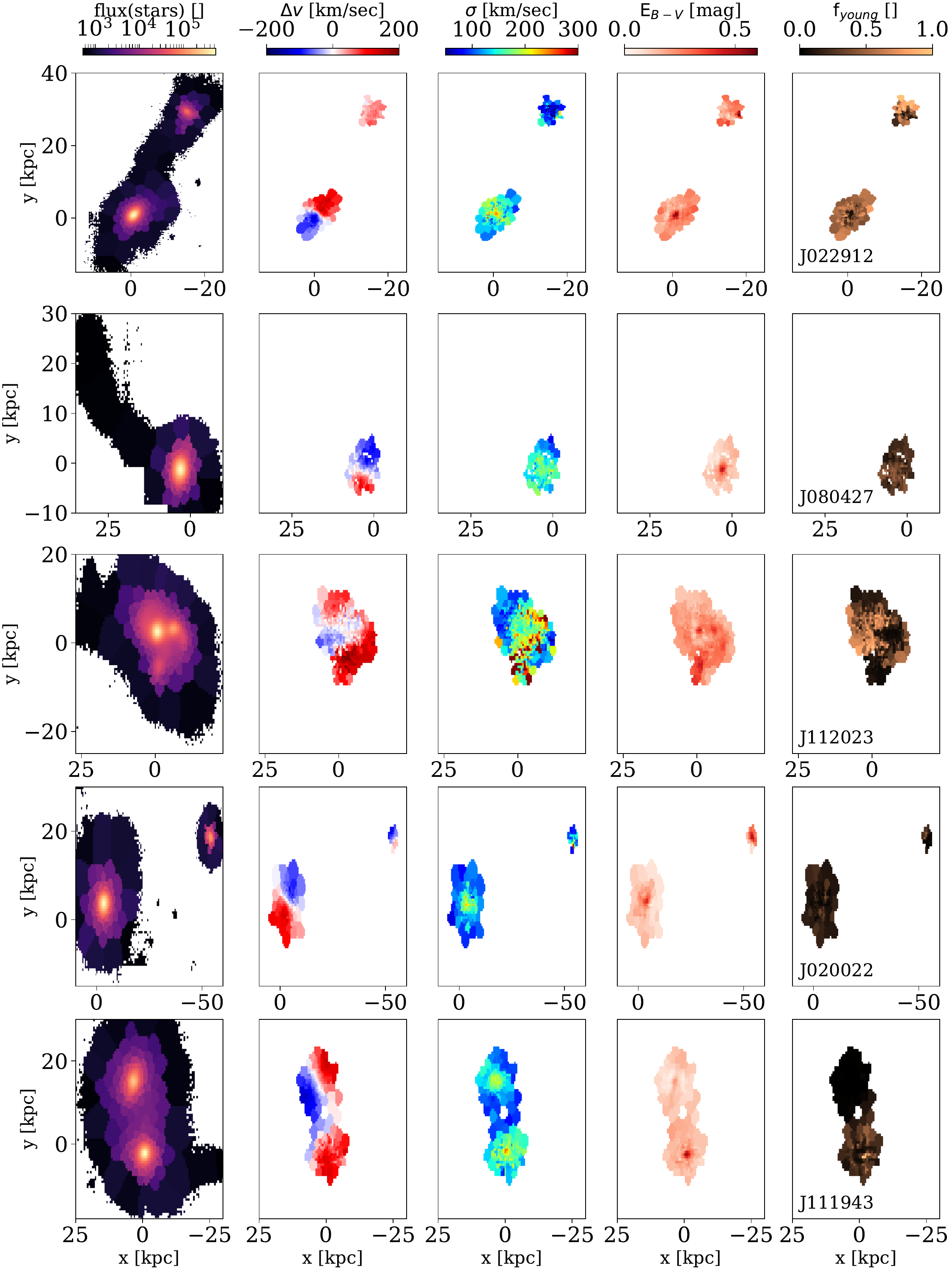}
\caption{\textbf{Spatially-resolved stellar properties.} Each row represents a galaxy in our MUSE sample. Within each row, the different panels show, from left to right: normalized stellar continuum emission in the range 5300--5700 \AA, stellar velocity with respect to systemic, stellar velocity dispersion, dust reddening towards the stars, and the fraction of young stars, which is defined as the sum of the weights of the stellar templates younger than 1 Gyr, divided by the sum of all the weights. 
}\label{f:stellar_parameters}
\end{figure*}

\subsection{Ongoing tidal interactions and/or mergers}\label{s:results:mergers}

The galaxies studied here were selected from our parent sample of post-starburst galaxies with AGN and evidence for ionized outflows \citep{baron22}. In particular, all the galaxies in our sample are H$\delta$-strong, with EW(H$\delta) > $ 5\AA, and with optical spectra that are dominated by A-type stars, suggesting a recent burst of star formation that was terminated abruptly $\sim$1 Gyr ago (e.g., \citealt{dressler83, couch87, poggianti99}). However, using IRAS 60 $\mathrm{\mu m}$ observations to estimate the SFR, in \citet{baron22} we showed that many systems selected as post-starbursts in the optical band host in fact obscured star formation, with some showing infrared luminosities comparable to local (U)LIRGs. We found that 45\% of galaxies in our parent sample are $>0.3$ dex above the star-forming main sequence (36\%--56\% at 95\% confidence), and 32\% are $>0.6$ dex above (24\%--41\%). For our combined sample of 7 galaxies observed with IFUs, table \ref{tab:gal_properties} shows that 6 galaxies have far-infrared luminosities $11 \leq \log L_{\mathrm{SF}}/L_{\odot} < 12$, while the remaining one has $\log L_{\mathrm{SF}}/L_{\odot} \approx 10.5$. These observations call into question the traditional interpretation of these sources as galaxies that started their transition to quiescence. In this section we use the spatially-resolved properties of the stars and gas in these galaxies to further address this question. The main properties we discuss are summarized in table \ref{tab:morphologies}.

In figure \ref{f:stellar_parameters} we show the spatially-resolved properties of the stars. In particular, we show the stellar continuum emission in the range 5300--5700 \AA, the stellar velocity and velocity dispersion, the reddening towards the stars, and the fraction of young stars. The latter is defined using the non-parametric SFH obtained with {\sc ppxf} as the sum of the weights of stellar templates younger than 1 Gyr, divided by the sum of all the weights. A fraction $f_{\mathrm{young}}=1$ represents spaxels where the spectrum is completely dominated by stars younger than 1 Gyrs, while $f_{\mathrm{young}}=0$ represents spaxels that are dominated by stars older than 1 Gyr. 

All five galaxies show evidence for a recent or ongoing interaction, with J022912, J020022, J111943 showing companion galaxies at the same redshift, J112023 showing at least two bright centers suggestive of a later-stage merger, and J080427 showing a tidal feature that extends to distances of $\sim$30 kpc. For the two previously-published post-starbursts, J003443 shows a companion galaxy \citep{baron18} and J124754 shows disturbed gas kinematics \citep{baron20}. Thus, among the 7 galaxies selected as post-starbursts, 6 show clear signatures for an early interaction or ongoing merger in their stellar continuum emission.

Interestingly, 3 out of the 5 galaxies presented here (and 4 out of the combined 7), are at an early stage of an interaction, with visible companions at distances $>10$ kpc. This suggests that post-starburst optical signatures may appear well before the final coalescence and starburst. It is consistent with the idea that these galaxies have already had their first close passage, leading to the increased star formation seen in far-infrared wavelengths, but with enough time that has passed so that some regions have already experienced rapid quenching and are traced by post-starburst signatures (e.g., \citealt{hopkins13}). 

Figure \ref{f:stellar_parameters} shows that 4 out of the 5 galaxies presented here (and 5 out of the combined 7) show ordered disk-like motions in their stellar kinematics, consistent with observations of other major mergers (e.g., \citealt{engel10, perna21}). The exception is the ongoing merger J112023 that shows non-ordered stellar velocities, very high velocity dispersions ($\sigma_{*} \sim$ 300 km/sec), and significant dust reddening at the outskirts of the galaxy. We do not find a strong correspondence between the fraction of young stars and the reddening towards the stars. Some spaxels show high fractions of young stars and high reddening values, as expected. However, other spaxels show high reddening values and little young stars. This may be due significant obscuration of the young stellar population in these spaxels. 

\begin{figure*}
	\centering
\includegraphics[width=0.85\textwidth]{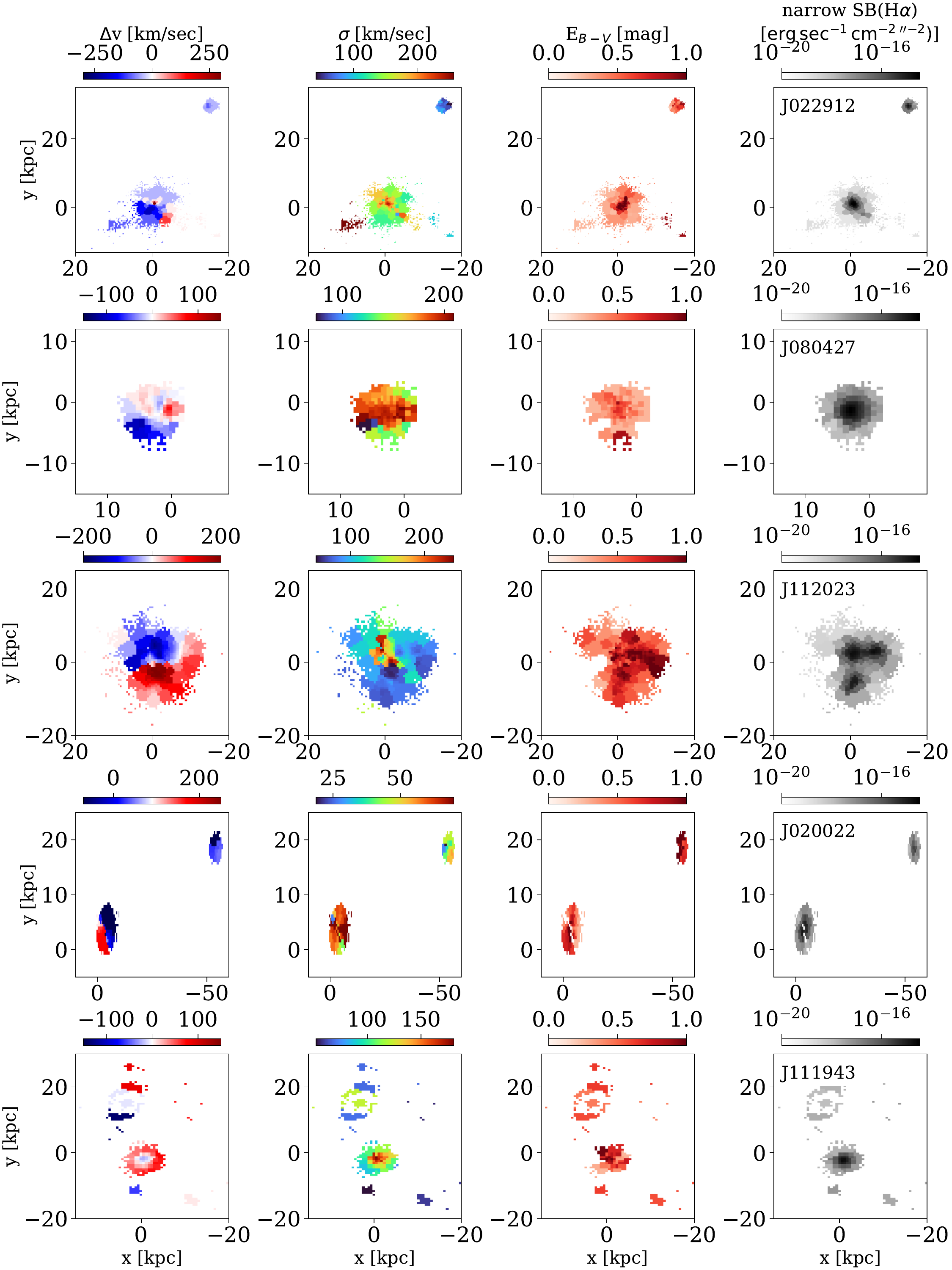}
\caption{\textbf{Spatially-resolved properties of the non-outflowing ionized gas.}  Each row represents a galaxy in our MUSE sample. Within each row, the different panels show, from left to right: the centroid velocity of the narrow kinematic component, the velocity dispersion, dust reddening towards the line-emitting region, and the surface brightness of the narrow H$\alpha$ flux. }\label{f:narrow_gas_properties}
\end{figure*}

In figure \ref{f:narrow_gas_properties} we show the spatially-resolved properties of the non-outflowing ionized gas that is traced by the narrow kinematic component. We show the gas velocity and velocity dispersion, reddening towards the line-emitting gas, and the surface brightness of the narrow H$\alpha$. In 3 out of the 5 galaxies (4 out of the combined 7) we find disturbed gas kinematics. The figure also shows significant reddening values, $\mathrm{E}(B-V)$ of 0.5--1 mag, in a large fraction of the spaxels. The reddening towards the narrow line-emitting gas is larger than towards the stellar continuum in most of the spaxels, which is in line with previous studies (e.g, \citealt{calzetti00, charlot00}).

\underline{To summarize:} all 7 galaxies from our combined sample show signatures of tidal interaction/merger in their morphology, stellar kinematics, or gas morphology and kinematics. The galaxies are at different stages of interaction, including a pair with a separation of $\sim$50 kpc with no visible tidal tails or bridges, an ongoing merger with two nuclei, and a single visible nucleus with a $\sim$30 kpc tidal feature. These observations suggest that post-starburst signatures in optical (i.e., strong H$\delta$ absorption) do not necessarily trace post-merger systems. Combined with our far-infrared and mm-based results \citep{baron22, baron23}, these observations suggest that systems selected as H$\delta$-strong with AGN and ionized outflows are more likely interacting dust-obscured starburst systems than their post-merger post-starburst descendants. Their far-infrared luminosities, $11 < \log L_{\mathrm{SF}}/L_{\odot} < 12$, are lower than those of $z < 0.15$ ULIRGs from the PUMA survey (which is designed to spatially-resolve the ionized gas in local ULIRGs; e.g., \citealt{perna21}). At this stage, it is not clear whether our galaxies are less luminous versions of the PUMA ULIRGs, or at an earlier stage of the evolution. We leave a thorough comparison between the samples to a future publication. 

\floattable
\begin{deluxetable}{c | m{5cm} | m{3cm} | m{2.3cm} | m{2.2cm} | m{3cm}}
\tablecaption{Broad emission line properties.\label{tab:broad_comps}}
\tablecolumns{6}
\tablenum{3}
\tablewidth{0pt}
\tablehead{
\colhead{Object ID} & \colhead{Red and blue wing geometry}  & \colhead{Reddening [mag]} & \colhead{Electron density [$\mathrm{cm^{-3}}$]}    & \colhead{Ionization}  & \colhead{Dominant origin}
}
\startdata
J022912 & \underline{Inner $\sim$5 kpc:} red and blue wings form double-cone-like geometry. The cone axis is perpendicular to the stellar rotation. & $\mathrm{E}_{B-V; n} = 0.90 \pm 0.30$ $\mathrm{E}_{B-V; b} = 0.86 \pm 0.34$ & $n_{e; n} = 130 \pm 100$ $n_{e; b} = 340 \pm 180$ & Seyfert-like, inconsistent with shocks  & nuclear (SN or AGN-driven) outflow \\
   & \underline{Outer regions:} low velocity dispersion, blue and red wings show comparable spatial extents, velocities, fluxes, and reddening. & $\mathrm{E}_{B-V; n} = 0.47 \pm 0.15$ $\mathrm{E}_{B-V; b} = 0.39 \pm 0.14$ & $n_{e; n} = 140 \pm 70$ $n_{e; b} = 320 \pm 150$ & LINER-like, consistent with shocks & non-nuclear flow at the galaxy edge (SN-driven winds or interaction-induced flows) \\
\hline
J080427   & Red and blue wings show comparable spatial extents, velocities, fluxes, and reddening. & $\mathrm{E}_{B-V; n} = 0.42 \pm 0.11$ $\mathrm{E}_{B-V; b} = 0.51 \pm 0.22$  & $n_{e; n} = 300 \pm 100$ $n_{e; b} = 40 \pm 20$ & Seyfert-like, inconsistent with shocks & non-nuclear flow at the galaxy edge (SN-driven winds or interaction-induced flows) \\
\hline
J112023  & \underline{Inner $\sim$5 kpc:} blue wing shows higher velocities and fluxes, red wing more reddened. & $\mathrm{E}_{B-V; n} = 0.70 \pm 0.13$ $\mathrm{E}_{B-V; b} = 0.33 \pm 0.20$ & $n_{e; n} = 150 \pm 70$ $n_{e; b} = 200 \pm 170$ & LINER-like, inconsistent with shocks & nuclear (SN or AGN-driven) outflow \\
   & \underline{Outer regions:} low velocity dispersion, blue and red wings show comparable spatial extents, velocities, fluxes, and reddening. & $\mathrm{E}_{B-V; n} = 0.87 \pm 0.12$ $\mathrm{E}_{B-V; b} = 0.36 \pm 0.15$  & $n_{e; n} = 50 \pm 30$ $n_{e; b} = 250 \pm 130$ & LINER and composite-like, consistent with shocks & interaction-induced flows  \\
\hline
J020022 & Red and blue wings form double-cone-like geometry. The cone axis is perpendicular to the stellar rotation. & $\mathrm{E}_{B-V; n} = 0.74 \pm 0.20$ $\mathrm{E}_{B-V; b} = 0.39 \pm 0.14$ & $n_{e; n} = 100 \pm 80$ $n_{e; b} = 140 \pm 80$ & consistent with shocks & interaction-induced flows (geometry inconsistent with SN-driven winds) \\
\hline
J111943 & Red and blue wings show comparable spatial extents, velocities, fluxes, and reddening. & $\mathrm{E}_{B-V; n} = 0.62 \pm 0.17$ $\mathrm{E}_{B-V; b} = 0.75 \pm 0.59$ & $n_{e; n} = 140 \pm 90$ $n_{e; b} = 640 \pm 460$ & inconsistent with shocks & non-nuclear flow \\
\enddata
\end{deluxetable}

\subsection{Broad wings of the ionized lines: outflows or interaction-induced motions?}\label{s:results:ionized}

Broad kinematic components are detected in the majority of spaxels in each of the galaxies in our sample. Visual inspection reveals high-SNR broad features, with peak fluxes that are 10--100\% of the peak fluxes of the narrow lines. In section \ref{s:results:mergers} we found that all the galaxies show signs of interactions/mergers, with some of the galaxies showing disturbed gas morphologies and kinematics. In such systems, broad kinematic components do not necessarily trace AGN or supernova-driven outflows, and may originate from galactic-scale gas flows caused by the interaction (see, for example, figure \ref{f:paper_cartoon}). In this section we study the properties of the ionized gas traced by the broad kinematic component, with the goal of constraining its origin. We summarize the different properties in table \ref{tab:broad_comps}, where we also list our suggested dominant origin for each of the components.

We derive the kinematics, spatial extent, fluxes and reddening, and dominant ionizing source for the blue and red wings separately, and use these, across many spaxels, to identify the most probable dynamical origin of the gas flows. In our classification, we distinguish between the case of a nuclear outflow (driven by the AGN, supernovae, or a combination of the two) and a non-nuclear flow that could be caused by supernovae at the closer edge of the galaxy or due to interaction-induced flows. Importantly, we make the distinction between the dynamical origin of the wind and the dominant ionization source of the gas in the wind. For example, it is possible that non-nuclear flows are primarily ionized by the AGN and show Seyfert-like optical line ratios.

\begin{figure*}
	\centering
\includegraphics[width=0.95\textwidth]{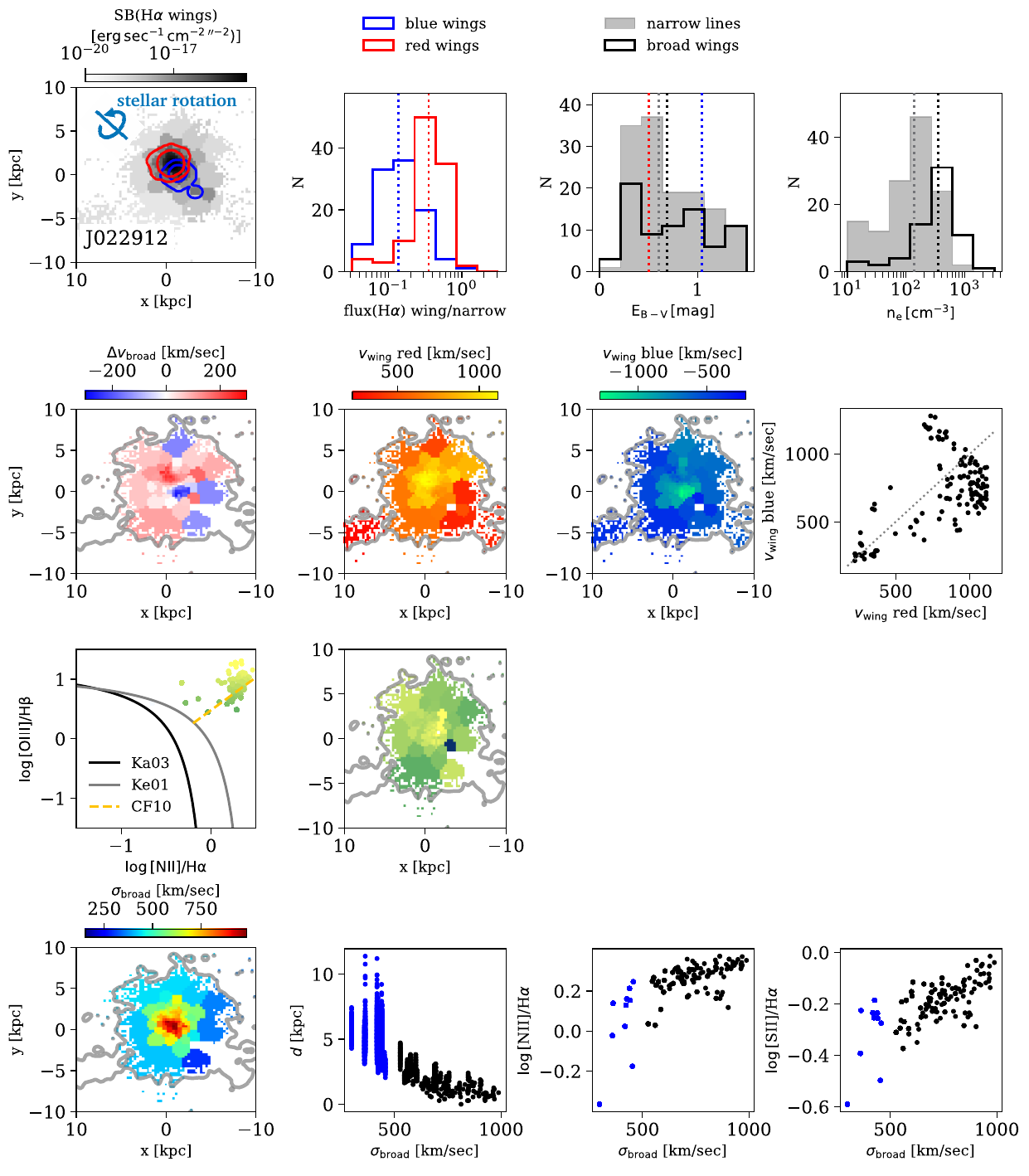}
\caption{\textbf{Broad line diagnostic diagrams for J022912.} Top row from left to right: (i) surface brightness of the broad H$\alpha$ wings (grey), with red and blue wings shown separately with contours, (ii) distribution of the flux in the wings, (iii) distribution of $\mathrm{E}_{B-V}$ measured towards the narrow lines (grey), broad wings (black), and red and blue wings, with dashed lines showing median values, and (iv) distribution of the electron density. Second row from left to right: (i) velocity of the broad kinematic component with respect to systemic, (ii) + (iii) maximum outflow velocity in the red and blue wings, and (iv) a comparison between the two. Third row: (i) the location of different spaxels on the line diagnostic diagram [NII]/H$\alpha$ versus [OIII]/H$\beta$ (\citealt{baldwin81, veilleux87}; separating criteria: \citealt{kewley01, kauff03a}; Ke01 and Ka03 respectively), (ii) and spatial distribution of the classification. Forth row: (i) velocity dispersion of the broad kinematic component, (ii) distance of each spaxel versus the velocity dispersion, and (iii) + (iv) the [NII]/H$\alpha$ and [SII]/H$\alpha$ broad line ratios versus the velocity dispersion.}\label{f:J022912_summary_plot}
\end{figure*}

For a nuclear outflow, we expect to find an asymmetry in either the kinematics, spatial extents, or fluxes and reddening between the red and the blue wing, when considering all the spaxels of a single galaxy. For example, a nuclear outflow might show a double-cone-like structure in kinematics and/or line extents (e.g., J022912 in figure \ref{f:J022912_summary_plot}), with the red wing dominating in one part of the galaxy and the blue wing in the other. For a nuclear wind viewed face-on, the red and blue wing kinematics and extents may be symmetric on galactic scales, but we expect the flux of the red wing to be lower than that of the blue wing, or to show significantly larger reddening values. This is because a larger fraction of the receding part of the outflow will be behind the stellar disc, compared to the approaching side of the flow (see figure \ref{f:paper_cartoon} and figure 14 in \citealt{davies_r14}). In case the red and blue wings show comparable velocities, symmetric extents on kpc scales, and comparable fluxes and reddening values (see e.g., J080427 in figure \ref{f:J080427_summary_plot}), we rule out the nuclear outflow origin. 

To study the dominant ionization source of the gas, we use the line ratios of the broad wings and their relation to the velocity dispersion. The galaxies in our sample were selected to have narrow emission line ratios consistent with AGN ionization (Seyfert or LINER) in their 1D SDSS spectrum. The spatially-resolved MUSE observations allow us to investigate the dominant ionization source in different regions within each galaxy, and study the contribution of shock excitation. We use the [OIII]/H$\beta$-[NII]/H$\alpha$ line diagnostic diagram (e.g., \citealt{baldwin81, veilleux87}) to classify the narrow and broad kinematic components separately. We use the separating criteria by \citet{kewley01}, \citet{kauff03a}, and \citet{cidfernandes10} to classify the spaxels into star-forming, composite, LINER, or Seyfert.

To check whether the broad line emission is consistent with shocks, we examine the relation between the line ratios [NII]/H$\alpha$ and [SII]/H$\alpha$ and the velocity dispersion of the broad kinematic component. Such relations have been found in star-forming galaxies and (U)LIRGs (e.g., \citealt{rich11, ho14, rich15, perna17}), and they indicate a coupling between the gas ionization and kinematics. Since such coupling in not predicted by photoionization models, studies attributed this relation to shock excitation. However, as discussed by \citet{laor98}, shocks are inefficient in converting mass to radiation, and as a result, even significant mass outflow rates will produce very little line radiation. Therefore, it is necessary to compare not only the line ratios, but also the line luminosities, to those predicted by shock models. We consider shock excitation as the dominant ionization mechanism if the following criteria are met: (i) there is a correlation between $\sigma_{\mathrm{broad}}$ and at least one of the line ratios [NII]/H$\alpha$ and [SII]/H$\alpha$, (ii) the observed line ratios are consistent with the predicted line ratios from fast radiative shocks by \citet{allen08} for densities in the range 10--1000 $\mathrm{cm}^{-3}$, and most importantly, (iii) the observed H$\alpha$ surface brightness is consistent, to a factor of 3, with the H$\alpha$ luminosity predicted by \citet{allen08}. As discussed later in this section, these criteria are met for the low velocity dispersion, low luminosity, extended broad lines in two galaxies. They are not met for the inner spaxels that show high velocity dispersions and high line luminosity.

\subsubsection{Discussion of individual sources}\label{s:results:ionized:individual_sources}

\underline{J022912:} figure \ref{f:J022912_summary_plot} summarizes the broad line diagnostic diagrams used in this section. Broad wings are detected throughout the FOV, and show similar extent to the narrow emission lines. The spaxels can be roughly divided into two classes, with the spaxels within $\sim$5 kpc showing high velocity dispersion of 500--1000 km/sec, high maximal outflow velocities of $>$800 km/sec, and Seyfert-like line ratios. Spaxels outside $\sim$5 kpc show lower velocities and velocity dispersions, and LINER-like line ratios. Within $\sim$5 kpc, the red and blue broad wings dominate different regions in the galaxy, forming a cone-like structure that is perpendicular to the rotating stellar disc. The maximum velocities in the red wings versus those of the blue wings (second row, right-most panel) form a relation that is perpendicular to the main diagonal, where the redshifted velocities are minimal when the blueshifted are maximal, and vice versa, supporting the double-cone outflow interpretation. In this region, the high line luminosities cannot be reproduced by shock models. Therefore, the properties of the broad wings within 5 kpc are consistent with a double-cone nuclear outflow (driven by the AGN, supernovae, or a combination of the two) that is ionized by the AGN. In the outer $>$5 kpc regions, the line ratios [NII]/H$\alpha$ and [SII]/H$\alpha$ increase with velocity dispersion, and the line luminosities can be reproduced by shock models. Therefore, we suggest that the broad wings at $>$ 5 kpc originate in a non-nuclear flow (supernova-driven winds at the edge of the galaxy or interaction-induced flows) that may be excited by shocks.

\underline{J080427:} figure \ref{f:J080427_summary_plot} in \ref{a:ionized_outflow_props} summarizes the broad line diagnostics. Broad wings are detected throughout the FOV with similar extent to that of the narrow lines. The red and blue wings show comparable extents, fluxes, and reddening values. This rules out a nuclear outflow viewed face on, where we expect the red wing to suffer from larger extinction than the blue wing. Moreover, the maximal velocities of the red/blue wings show a trend along the main diagonal, suggesting that the redshifted velocities are maximal where the blueshifted are maximal. While all these observables may be consistent with a double-cone outflow viewed edge-on, we rule out this option since the flux in the red/blue wings shows spherical morphology. Therefore, in this case, we favor the non-nuclear flow interpretation (supernovae at the edge of the galaxy and/or merger-induced flows). The line ratios are consistent with AGN-SF composite and with Seyfert ionization. We rule out shock excitation since [NII]/H$\alpha$ and [SII]/H$\alpha$ do not increase with increasing velocity dispersion, and since shock models cannot reproduce the observed line luminosity in most of the spaxels. 

\underline{J112023:} figure \ref{f:J112023_summary_plot} in \ref{a:ionized_outflow_props} summarizes the broad line diagnostics. Broad wings are detected throughout a large fraction of the FOV and show roughly similar extent to that of the narrow lines. Similarly to J022912, the spaxels can be divided into two classes. Spaxels in the inner $\sim$5 kpc of the primary galaxy show high velocity dispersions of $>$400 km/sec and high maximal outflow velocities of $>$500 km/sec. The line ratios are consistent with LINER ionization, but show no relation to the velocity dispersion, ruling out shock excitation. The broad red and blue wings show asymmetry in the maximum outflow velocity, with the blue wing showing higher outflow velocities in all the spaxels. These observations are consistent with a spherical or double-cone outflow viewed face on, with the receding part of the flow suffering from large extinction. We therefore suggest that the broad wings in the inner $\sim$ 5 kpc originate from a nuclear outflow that is ionized by the central AGN. The outer spaxels show low velocity dispersions of $\sim$200--300 km/sec. The line ratios are consistent with AGN-SF composite and with LINER, and increase with increasing velocity dispersion. Both the line ratios and luminosities can be reproduced with shock models. The low velocities and reddening values favor merger-induced motions as the origin of these broad features.

\underline{J020022:} figure \ref{f:J020022_summary_plot} in \ref{a:ionized_outflow_props} summarizes the broad line diagnostics. The broad wings show similar extent to that of the narrow lines, and show a clear cone-like structure. However, the double-cone is parallel to the stellar disc rotation,  with the blueshifted part of the cone corresponding to blueshifted stellar velocities and vice versa for the red. This is inconsistent with the orientation of AGN or supernova-driven outflows in simulations (e.g., \citealt{nelson19}), and may suggest that the broad component originates from interaction-induced flows. For J020022, the [OIII] is masked-out by the reduction pipeline since it coincides with the WFM-AO laser’s sodium lines. We therefore use the integrated [OIII]/H$\beta$ line ratio from the SDSS, which places the spaxels in the Seyfert region in the BPT diagram. The observed [NII]/H$\alpha$ line ratio increases with velocity dispersion, and shock models can reproduce the observed line luminosity. 

\underline{J111943:} figure \ref{f:J111943_summary_plot} in \ref{a:ionized_outflow_props} summarizes the broad line diagnostics. The broad wings show similar extents to that of the narrow lines, with the red and blue wings having similar extents and comparable flux and reddening values. This rules out nuclear outflow origin for this source. Similarly to J020022, the [OIII] is masked-out by the reduction pipeline and we use the integrated [OIII]/H$\beta$ line ratio from the SDSS. The line ratios are consistent with either LINER or Seyfert radiation, and they show no correlation with the velocity dispersion. We therefore rule out shock excitation as the main source of ionizing radiation.

\begin{figure*}
	\centering
\includegraphics[width=0.8\textwidth]{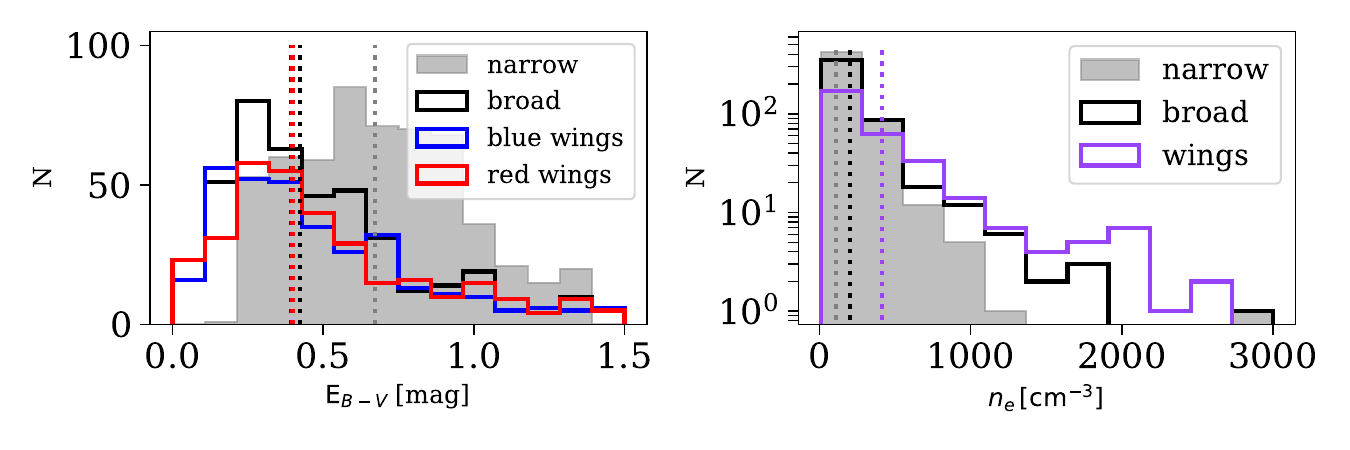}
\caption{\textbf{Reddening and electron density distributions for different kinematic components.} \textbf{The left panel} compares the distributions of reddening measured using the narrow lines (grey), broad lines (black), and red and blue wings (red and blue respectively), in every Voronoi bin, for all the galaxies we consider. The dashed lines represent the median values. Similarly to other studies, when considering all bins from all galaxies together, the reddening for the broad lines and red/blue wings are lower than those of the narrow lines. \textbf{The right panel} compares the distribution of electron density measured for the narrow lines (grey), broad lines (black), and red+blue wings (purple), using the [SII] doublet, in every Voronoi bin and for all the galaxies we consider. Similarly to other studies, we find higher electron densities in the broad lines/ wings. This simple picture of lower reddening and higher electron density of the outflow can only be reproduced when considering all the spaxels from all the galaxies combined. Examining each source individually reveals a significant diversity in reddening and density properties. These properties can be used to place constraints on the outflow origin, and considering only the combined distributions may average out important differences.}\label{f:reddening_and_ne_comparison_all}
\end{figure*}

\subsubsection{General properties of the sample}\label{s:results:ionized:general_conclusions}

In figure \ref{f:reddening_and_ne_comparison_all} we show the distribution of reddening and electron density values for different kinematic components (narrow, broad, red or blue wings, red + blue wings) across all the Voronoi bins of the different galaxies. We find comparable reddening values for the broad lines and red/blue wings, and find them to be significantly lower than the reddening values derived using the narrow lines. This is in line with the results by \citet{mingozzi19} and \citet{fluetsch21} for local AGN and (U)LIRGs respectively, where the studies classified all broad components as originating from an outflow. However, the individual diagnostic diagrams (figures \ref{f:J022912_summary_plot}, \ref{f:J080427_summary_plot}, \ref{f:J112023_summary_plot}, \ref{f:J020022_summary_plot}, and \ref{f:J111943_summary_plot}) show a diversity in the reddening properties, with some spaxels showing significantly-higher reddening values in their broad components and wings compared to the narrow lines. This diversity in reddening can be used to place additional constraints on the outflow geometry and its origin, and by considering only the combined distributions from all the sources, important differences may be averaged out.

The right panel of figure \ref{f:reddening_and_ne_comparison_all} compares between the electron densities derived for the narrow lines, broad lines, and red+blue wings, using the [SII] doublet. When considering all the bins, we find that the electron density of the wings ($\sim$300 $\mathrm{cm^{-3}}$) is higher than the electron density of the broad component ($\sim$200 $\mathrm{cm^{-3}}$), which is higher than that of the narrow component ($\sim$100 $\mathrm{cm^{-3}}$). This too, is in line with the results by \citet{mingozzi19} and \citet{fluetsch21}. However, inspecting the individual diagnostic diagrams reveals that this is not the case for J080427, where the electron density in the wings/broad component is lower than that of the narrow lines. 

We used the ionization parameter method to derive the electron density in spaxels that are dominated by AGN photoionization (\citealt{baron19b}; see table \ref{tab:ionized_outflow_properties}). The resulting electron densities are comparable to those derived using the [SII] for two sources (J022912 and J020022), and are 10--100 times larger than the [SII]-based ones for three galaxies (J080427, J112023, and J111943). For the estimates of mass and energetics of the ionized outflows, we use the [SII]-based estimates to allow for a straightforward comparison with other studies, and since the ionization parameter method can only be applied to spaxels dominated by AGN photoionization. 

\underline{To summarize:} we detect broad kinematic components in the majority of the spaxels in each of the primary galaxies. These components show diverse and complex kinematics, flux extents, reddening values, electron densities, and ionization mechanisms. This diversity suggests that a general comparison between narrow (=disk) and broad (=outflow) kinematic components may average out important differences between galaxies whose broad lines may originate from different physical processes (e.g., supernova-driven outflows, AGN-driven outflows, and galactic-scale flows due to interactions). Our detailed analysis of the line properties revealed that the observed broad emission lines are inconsistent with AGN-driven outflows in 3 out of 5 galaxies. This is in stark contrast to our initial expectation and the default assumption by most studies that high-velocity ($v > 500$ km/sec) ionized gas in active galaxies originates from AGN-driven outflows.

\begin{figure*}
	\centering
\includegraphics[width=1\textwidth]{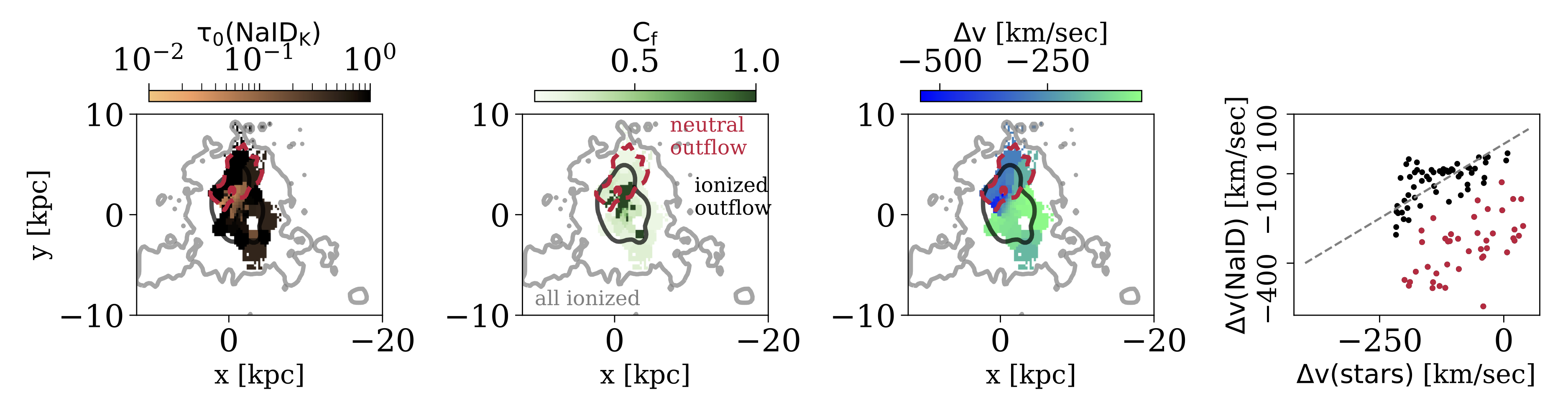}
\caption{\textbf{Spatially-resolved properties of the NaID absorption in J022912.} From left to right: the best-fitting absorption optical depth of the NaID$_{\mathrm{k}}$ doublet component, the best-fitting gas covering factor, the centroid velocity of the NaID absorption with respect to systemic, and a comparison between the centroid velocity of the NaID absorption and the stars. Blueshifted NaID absorption is considered as an outflow if it is blueshifted with respect to the stellar velocity by more than 100 km/sec. }\label{f:J022912_NaID_abs_properties_edited}
\end{figure*}

\begin{figure*}
	\centering
\includegraphics[width=1\textwidth]{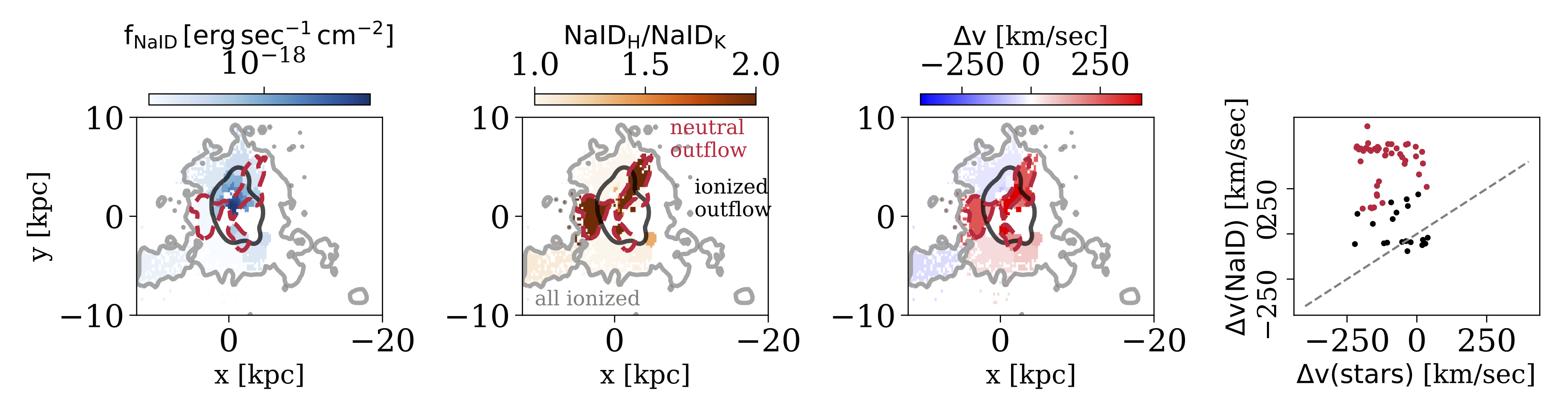}
\caption{\textbf{Spatially-resolved properties of the NaID emission in J022912.} From left to right: the integrated flux of the NaID emission, the amplitude ratio of the $H$ and $K$ components, where $\mathrm{NaID}_{H}/\mathrm{NaID}_{K}=1$ for optically-thick gas, and $\mathrm{NaID}_{H}/\mathrm{NaID}_{K}=2$ for optically-thin gas, the centroid velocity of the NaID emission with respect to systemic, and a comparison between the centroid velocity of the NaID emission and the stars. Redshifted NaID emission is considered as an outflow if it is redshifted with respect to the stellar velocity by more than 100 km/sec.  }\label{f:J022912_NaID_emis_properties_edited}
\end{figure*}

\begin{figure}
\includegraphics[width=3.5in]{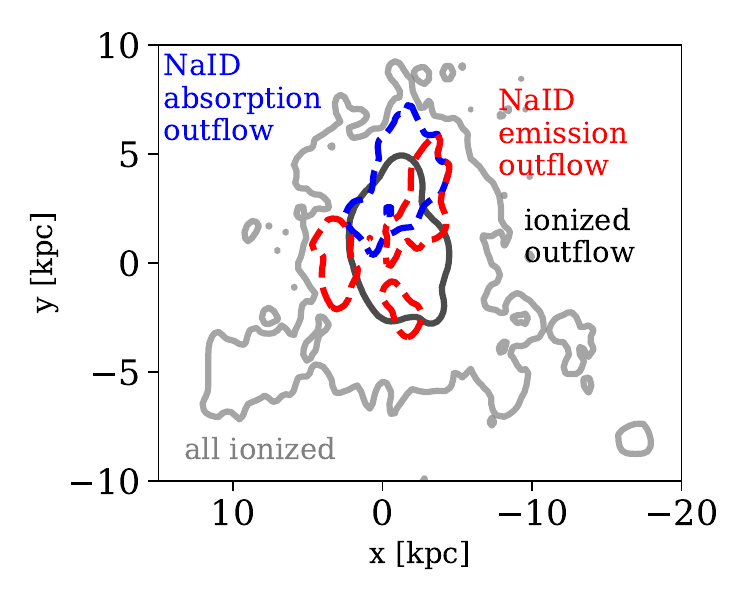}
\caption{\textbf{Neutral and ionized outflow extents in J022912.} The figure shows the distribution of the narrow line-emitting gas (grey contours), the nuclear ionized outflow (black contours), the blueshifted NaID absorption (blue contours) and redshifted emission (red contours) that are associated with the neutral outflow. The blueshifted NaID absorption and redshifted NaID emission show up in distinct regions in the galaxy.} 
\label{f:J022912_NaID_emis_and_abs_rep}
\end{figure}

\subsection{Complex picture of the neutral gas as traced by NaID absorption and emission}\label{s:results:neutral}

In section \ref{s:results:ionized} above we performed a detailed analysis of the broad wings of the ionized lines, finding them to be inconsistent with AGN-driven outflows in 3 out of 5 of the cases. Furthermore, in some galaxies, the observations suggested that the broad emission lines originate from galactic-scale interaction-induced motions, rather than from outflows. These observations raise the question of how to interpret broad kinematic emission components, given that both inflows and outflows can produce redshifted and blueshifted wings. Contrary to the ionized lines, the resonant NaID absorption does not suffer from the inflow-outflow degeneracy. In the case of an outflow, the NaID profile will resemble a P-Cygni profile, with blueshifted absorption and redshifted emission. In the case of an inflow, the profile is expected to be reversed, showing redshifted absorption and blueshifted emission. Therefore, NaID emission and absorption may be more straightforward to interpret in interacting and/or merging systems, in particular in cases where spatial information (e.g., IFU) is not available. 

Table \ref{tab:morphologies} summarizes the NaID properties of the galaxies in our sample. NaID absorption is detected in the three galaxies: J022912, J080427, and J112023. Interestingly, NaID emission is detected in 2 galaxies, which are the only two galaxies for which a nuclear outflow origin (either AGN-driven wind or central supernova-driven wind) is our favored interpretation for the broad ionized lines. NaID was neither detected in absorption nor in emission for J020022 and J111943. For these two galaxies, the observations suggested that the broad ionized lines originate from either interaction-induced flows or supernova-driven winds at the closer edge of the galaxy. 

We now focus on the NaID profile of J022912, which is the most complex out of the three. A similar analysis has been applied to J080427 and J112023 (see figure \ref{f:neutral_outflow_properties}), and we report the mass and energetics of the neutral outflows for all three galaxies in section \ref{s:results:outflows}. In figure \ref{f:J022912_NaID_abs_properties_edited} we show the spatially-resolved properties of the NaID absorption for J022912. In particular, we show the best-fitting absorption optical depth, covering factor, and centroid velocity with respect to systemic. The observations suggest optically-thin absorption, with a unity covering factor in the center of the galaxy and low covering factors outside. The spatially-resolved centroid velocity is consistent with an outflowing gas cone with a large opening angle. The figure also compares between the centroid velocity of the NaID absorption and the stars. Our conservative approach is to consider the blueshifted NaID absorption as an outflow only if it is blueshifted with respect to the stars by more than 100 km/sec, which is indicated in the diagram.

In figure \ref{f:J022912_NaID_emis_properties_edited} we show the spatially-resolved properties of the NaID emission for J022912. We show the derived NaID emission flux, the doublet amplitude ratio $\mathrm{NaID}_{H}/\mathrm{NaID}_{K}$, where a ratio of 1 (2) suggests optically-thick (optically-thin) gas, and the centroid velocity of the NaID emission with respect to systemic. Contrary to the NaID absorption that shows only blueshifted velocities, the NaID emission shows both blueshifted and redshifted NaID emission. Since this gas is located at the farther side of the galaxy, behind the stars, the blueshifted emission can be easily interpreted as an inflow and the redshifted emission as an outflow. These two components show distinct $\mathrm{NaID}_{H}/\mathrm{NaID}_{K}$ ratios, where the inflow is optically-thick and the outflow is optically-thin. Similarly to the absorption case, we consider redshifted NaID emission as an outflow only if it is redshifted with respect to the stars by more than 100 km/sec. 

In figure \ref{f:J022912_NaID_emis_and_abs_rep} we compare between the extents of the narrow line-emitting gas, the nuclear ionized outflow, and the blueshifted NaID absorption and redshifted emission that are associated with the neutral outflow. The extent of the neutral outflow is comparable to that of the nuclear ionized outflow. Interestingly, the blueshifted NaID absorption and redshifted NaID emission are most dominant in different regions in the galaxy. This seems to be a generic property of neutral outflows, where similar behavior is seen for J112023 (see figure \ref{f:neutral_outflow_properties}), J124754 (\citealt{baron20}), and F05189-2524 (\citealt{rupke15}). It may be partially due to absorption-emission line filling and velocity projections. It further emphasizes the importance of taking into account the NaID emission when estimating the mass and energetics of neutral outflows, as it traces separate regions of the outflow.  

We do not find a significant connection between the neutral and ionized outflow phases (see figure \ref{f:J022912_neutral_to_ionized_relation} in \ref{a:NaID_fitting} for J022912). In particular, we do not find a significant relation between the redshifted NaID velocity and the velocity of the red H$\alpha$ wing, and similarly for the blueshifted NaID absorption and the blue H$\alpha$ wing. In addition, we do not find a significant relation between the NaID EW or flux and the flux of the H$\alpha$ wings. Nevertheless, similarly to J124754 (\citealt{baron20}), we find that the NaID emission to H$\alpha$ flux ratio is about 0.1. 

\floattable
\begin{deluxetable}{c c c c c c c c c c}
\tablecaption{Ionized outflow properties\label{tab:ionized_outflow_properties}}
\tablecolumns{10}
\tablenum{4}
\tablewidth{0pt}
\tablehead{
\colhead{(1)}      & \colhead{(2)} & \colhead{(3)} & \colhead{(4)}   & \colhead{(5)} & \colhead{(6)} & \colhead{(7)} & \colhead{(8)}  & \colhead{(9)}  & \colhead{(10)}\\
\colhead{Object ID} & \colhead{$v_{\mathrm{out}}$} & \colhead{$r_{95\%}$} & \colhead{$r_{\mathrm{avg}}$} & \colhead{$\log n_{\mathrm{e}}$([SII])} & \colhead{$\log U$} & \colhead{$\log n_{\mathrm{e}}$($\log U$)} & \colhead{$\log M_{\mathrm{ion}}$} & \colhead{$\log \dot{M}_{\mathrm{ion}}$} & \colhead{$\log \dot{E}_{\mathrm{ion}}$} \\
\colhead{} & \colhead{[km/sec]} & \colhead{[kpc]} & \colhead{[kpc]} & \colhead{[$\mathrm{cm^{-3}}]$} & \colhead{[]} & \colhead{[$\mathrm{cm^{-3}}]$} & \colhead{[$M_{\odot}$]} & \colhead{[$M_{\odot}$/yr]} & \colhead{[erg/sec]}
}
\startdata
J022912               & 1000 & 4.5 & 4.0 & 2.6 & -3.3 & 2.6 & 7.61 & 0.98  & 42.48 \\
J080427               & 700  & 5.0 & 1.0 & 1.7 & -3.3 & 3.6 & 8.28 & 1.44  & 42.64 \\
J112023               & 800  & 5.0 & 2.5 & 2.5 & -3.7 & 3.5 & 7.30 & 0.52  & 41.83 \\
J020022$^{\dagger}$   & 400  & 1.7 & 1.4 & 2.1 & -3.0 & 2.0 & 6.60 & 0.02  & 40.69 \\
J111943               & 500  & 2.3 & 0.8 & 3.0 & -3.4 & 3.9 & 7.45 & 0.81  & 41.71 \\
\enddata
\tablecomments{ \footnotesize{\textbf{Columns.} (1): Object identifier. (2) Adopted velocity of the ionized outflow. (3) The adopted extent of the outflow, which is the 95th percentile of the distance distribution of individual spaxels where an outflow have been identified. (4) Average extent of the outflow, obtained by an H$\alpha$ flux-weighted average of the distance of individual spaxels. (5) Electron density in the outflow using the [SII] method. (6) Ionization parameter of the AGN-ionized gas using the \citet{baron19b} method. (7) Electron density in the outflow using the ionization parameter method. (8) Outflowing ionized gas mass. (9) Mass outflow rate of the ionized gas. (10) Kinetic power of the ionized gas. \\
\textbf{Notes.} $^{\dagger}$: according to our interpretation, the broad emission lines in this galaxy originate from galactic-scale flows caused by the interaction and not by outflows.
}}
\end{deluxetable}

\floattable
\begin{deluxetable}{c r r r r r r}
\tablecaption{Neutral outflow properties\label{tab:neutral_outflow_properties}}
\tablecolumns{7}
\tablenum{5}
\tablewidth{0pt}
\tablehead{
\colhead{(1)}      & \colhead{(2)} & \colhead{(3)} & \colhead{(4)}   & \colhead{(5)} & \colhead{(6)} & \colhead{(7)} \\
\colhead{Object ID} & \colhead{$\log M_{\mathrm{abs}}$} & \colhead{$\log \dot{M}_{\mathrm{abs}}$} & \colhead{$\log \dot{E}_{\mathrm{abs}}$} & \colhead{$\log M_{\mathrm{emis}}$} & \colhead{$\log \dot{M}_{\mathrm{emis}}$} & \colhead{$\log \dot{E}_{\mathrm{emis}}$} \\
\colhead{} & \colhead{[$M_{\odot}$]} & \colhead{[$M_{\odot}$/yr]} & \colhead{[erg/sec]} & \colhead{[$M_{\odot}$]} & \colhead{[$M_{\odot}$/yr]} & \colhead{[erg/sec]}  \\
}
\startdata
J022912   & 7.54     & 0.67 & 42.01 & 8.01    & 0.37 & 41.36 \\
J080427   & 7.34     & 0.28 & 41.09 &         &      &       \\
J112023   & 8.72     & 1.22 & 42.50 & 9.05    & 0.70 & 41.72 \\
J020022$^{\dagger}$   & $<10.06$ &      &       & $<7.71$ &      &       \\
J111943   & $<7.09$  &      &       & $<7.99$ &      &       \\
\enddata
\tablecomments{ \footnotesize{\textbf{Columns.} (1): Object identifier. (2) Outflowing gas mass of the neutral gas that is traced by NaID absorption. (3) Mass outflow rate of the neutral gas that is traced by NaID absorption. (4) Kinetic power of the neutral gas that is traced by NaID absorption. (5) Outflowing gas mass of the neutral gas that is traced by NaID emission. (6) Mass outflow rate of the neutral gas that is traced by NaID emission. (7) Kinetic power of the neutral gas that is traced by NaID emission.\\
\textbf{Notes.} $^{\dagger}$: according to our interpretation, the broad emission lines in this galaxy originate from galactic-scale flows caused by the interaction and not by outflows.
}}
\end{deluxetable}

\subsection{Multiphased wind energetics}\label{s:results:outflows}

In tables \ref{tab:ionized_outflow_properties} and \ref{tab:neutral_outflow_properties} we list the derived mass and energetics of the ionized and neutral winds for the galaxies in our sample. As noted in sections \ref{s:results:ionized} and \ref{s:results:neutral}, the derived red and blue wing kinematics, spatial extents, fluxes and reddening, for the ionized lines are consistent with a nuclear outflow origin (AGN, supernovae, or a combination of the two), in 2 out of the 5 galaxies. These galaxies, J022912 and J112023, are also the only two systems for which NaID has been detected in emission. These two galaxies have AGN bolometric luminosities of $\log L_{\mathrm{bol}}=45$. For these galaxies, the mass outflow rates of the ionized gas are 10 and 3 $M_{\odot}$/yr respectively, compared to the mass outflow rate of the neutral outflow which is 7 and 21 $M_{\odot}$/yr respectively. Assuming that the multiphased outflows are driven solely by the AGN, the wind coupling efficiency is estimated to be $\sim$0.5\%. These systems are also luminous infrared galaxies with the highest star formation luminosities within our sample ($\log L_{\mathrm{SF}}$ of 11.3 and 11.5 $L_{\odot}$, equivalent to SFRs of 20 and 30 $M_{\odot}/\mathrm{yr}$, respectively). In such systems, the observed winds are probably driven by a combination of supernovae and AGN feedback (e.g., \citealt{nelson19}), making the derived coupling efficiencies upper limits. 

\begin{figure*}
	\centering
\includegraphics[width=0.95\textwidth]{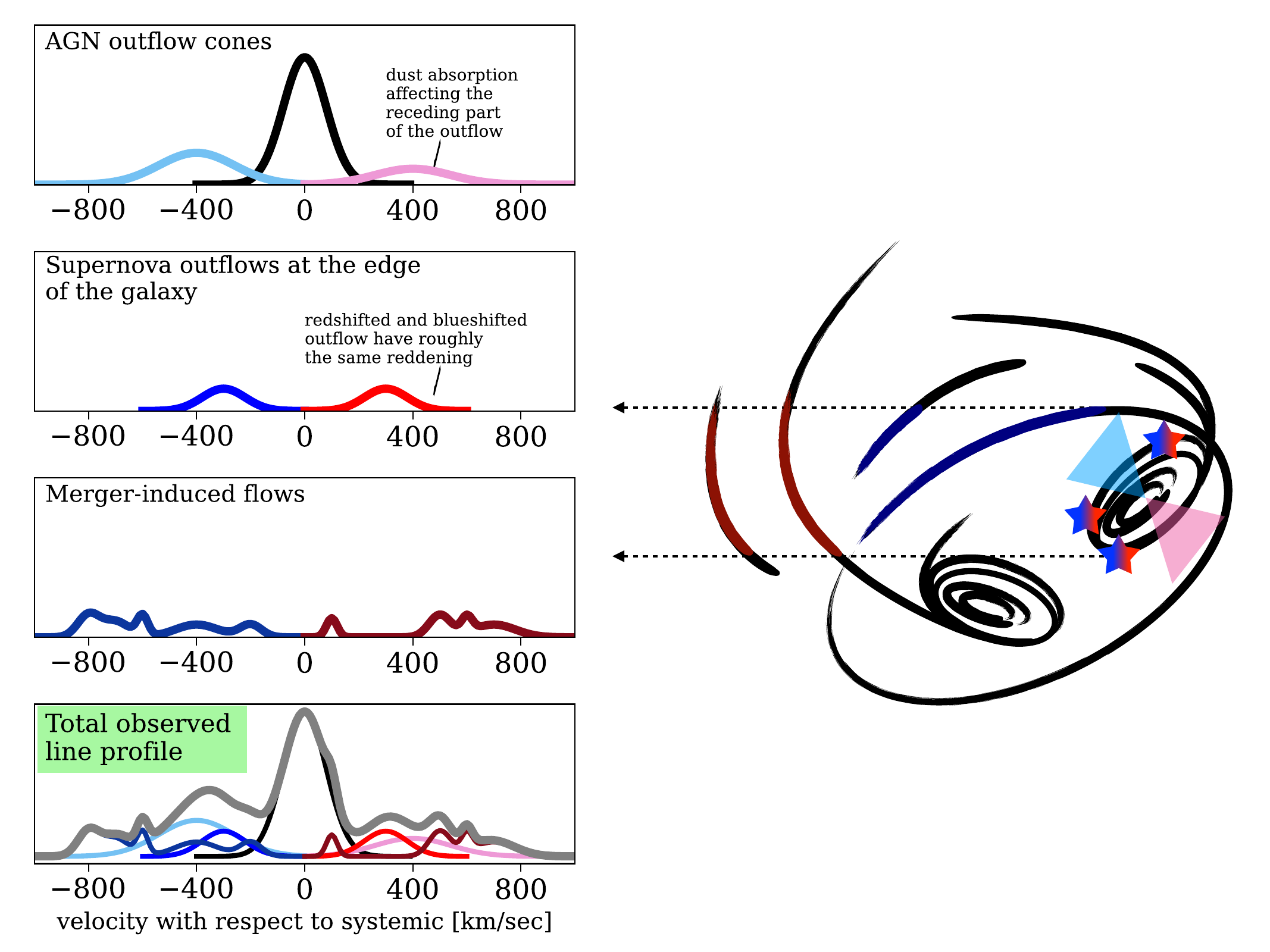}
\caption{\textbf{Emerging picture of the warm ionized gas in interacting galaxies.} The cartoon on the right depicts two interacting galaxies, where gas is stripped during the interaction to form tidal features. The upper galaxy hosts AGN-driven winds in a cone-like structure and supernovae-driven winds shown as stars. The field of view is marked with the horizontal dashed lines. The left panels represent the warm ionized gas line emission of the different components, showing that the total line emission is a combination of emission from the stationary gas in the primary galaxy, nuclear outflows due to AGN and/or supernovae, non-nuclear supernovae-driven winds (e.g., at the edge of the galaxy), and merger-induced gas flows. The line fluxes and gas velocity scales were arbitrarily chosen and are not based on a physically-motivated model of the expected gas kinematics of the different processes. The cartoon illustrates that the blueshifted and redshifted wings that are typically attributed to AGN winds can be in fact a complex combination of multiple dynamical processes. Spatially-resolved observations of multiple lines tracing the flow may be used to place constraints on the dominant dynamical process.}\label{f:paper_cartoon}
\end{figure*}

Table \ref{tab:neutral_outflow_properties} suggests that the mass and energetics of the neutral outflow as traced by NaID emission are comparable to those traced my NaID absorption. Therefore, we suggest that, when detected, NaID emission should be taken into account when estimating the mass and energetics of neutral outflows. The derived mass and energetics of the neutral outflows are within the ranges observed in (U)LIRGs+AGN by \citet{rupke05b} and by \citet{fluetsch21} for AGN of comparable luminosities. Similarly, the derived mass and energetics of the ionized outflows are comparable to those reported by \citet{fluetsch21} for (U)LIRGs+AGN with comparable luminosities.

\section{Discussion}\label{s:discussion} 

Figure \ref{f:paper_cartoon} depicts the emerging picture of the warm ionized gas in the sources in our sample. The observed broad blueshifted and redshifted wings may be a complex combination of nuclear outflows due to AGN and/or supernovae, non-nuclear supernovae-driven outflows (e.g., at the edge of the galaxy), and merger-induced flows. The gas velocities and line fluxes were arbitrarily chosen and are not based on a physically-motivated model for the gas kinematics expected from each of the dynamical processes. However, simulations of galaxy major mergers find that the warm ionized gas in merger-induced flows can reach velocities of 500--750 km/sec on tens of kpc scales (e.g., \citealt{hopkins13b}). The cartoon illustrates the challenge in disentangling the different processes even with spatially-resolved observations. Importantly, a distinction should be made between the process that ionizes/excites the gas and the dynamical origin of the flow. For example, tidally-stripped gas may be primarily ionized by the AGN, showing significant blueshifted and redshifted emission lines with Seyfert-like line ratios. Therefore, line ratios alone cannot be used to distinguish between different processes and isolate the contribution of AGN feedback. In this paper we used the spatial distribution of the flux in the redshifted and blueshifted wings, the derived reddening, and kinematics to place constraints on the dynamical origin of the observed flows. 

The sample presented in this paper was selected from our parent sample of post-starburst galaxies with AGN and indications for an ionized outflow \citep{baron22}. In our selection, we assumed that systems that show a combination of narrow and broad kinematic components in their recombination and forbidden lines host ionized outflows. Therefore, we formed our parent sample by selecting galaxies that show broad kinematic components in their optical emission lines. This is a typical assumption and choice in studies of outflows in active galaxies, in particular in cases where only a small subset of objects can be followed-up with IFU observations (e.g., \citealt{mullaney13, harrison14, bae17, fiore17, mingozzi19, fluetsch21}). Despite this selection, our analysis suggests that the broad emission lines originate from nuclear outflows (AGN or supernovae driven) only in 2 out of 5 objects. In the rest, the broad emission line properties are more consistent with a non-nuclear dynamical origin, for example, interaction-induced galactic-scale flows. This suggests that selecting galaxies with broad emission lines may bias the sample to include a larger fraction of interacting systems. Estimates of SFR, stellar mass, and morphological class can be used to remove interacting galaxies, thus minimize this bias.

While this does not pose a particular problem for studies of AGN in low-z and high-z galaxies that show no dynamical disturbance (e.g, low-z: \citealt{davies_r14, mingozzi19}; high-z: \citealt{genzel14, forster_schreiber19}), it may be a more significant challenge for infrared-bright galaxies, which are more likely to be interacting systems. In such systems, the close interaction may produce high gas velocities on galactic scales, which appear as broadened kinematic components in the emission lines and may mistakenly be classified as galactic-scale outflows. In particular, the largest compilations of multi-phased outflows in active galaxies (\citealt{fiore17, fluetsch19, fluetsch21}) include a large fraction of infrared-bright galaxies in which the outflows have been detected primarily in emission (in optical or mm wavelengths), and it is not clear what is the contribution of merger-induced motions to the observed broadened lines. In such cases, absorption lines may be more straight-forward to interpret as outflows (e.g., \citealt{rupke05a, rupke17}). 

Since quasar activity is linked to galaxy interactions (e.g., \citealt{sanders96, genzel98, hopkins06, veilleux09}; and more recently, e.g., \citealt{dougherty23, hernandez_toledo_23, pierce23}), our results raise the question of how to interpret broadened kinematic components observed in quasar spectra, in particular where spatially-resolved information is not available, and where the Balmer and the lower ionization [NII] and [SII] lines that may trace the outflow are often blended with the broad H$\alpha$ and H$\beta$ lines originating in the BLR. The detection of a blueshifted broad wing without a redshifted broad wing in a large sample of sources may indicate an outflow origin, rather than merger-induced flows, on average (e.g., \citealt{zakamska14, perna15, zakamska16}). However, a non-negligible fraction of these quasars show evidence for a redshifted broad wing as well (see e.g., figure 2 in \citealt{zakamska14}). Since the red and blue broadened wings of the H$\alpha$ and H$\beta$ lines are blended with the BLR H$\alpha$ and H$\beta$, one cannot derive the reddening of the kinematic components to test whether these cases are consistent with a nuclear outflow origin.

\section{Summary and Conclusions}\label{s:conclusions} 

Post-starburst E+A galaxies are believed to be the evolutionary link between major merger (ultra)luminous infrared galaxies and quenched ellipticals. Both observations and simulations suggest that this transition is rapid, with the starburst quenching abruptly over a timescale of a few hundreds Myrs. Although simulations invoke AGN feedback as one of the processes responsible for the rapid quenching of star formation, little is known observationally about AGN feedback, in particular AGN-driven winds, in this stage. To study the role of AGN feedback in the transition from starburst to quiescence, we constructed a sample of galaxies with post-starburst signatures in optical (strong H$\delta$ absorption), evidence for an AGN (using narrow line ratios), and evidence for ionized outflows (presence of broad kinematic components in H$\alpha$ and [OIII]). We presented the full sample in \citet{baron22}, where we found that a large fraction of the post-starburst galaxies host obscured star formation, with some systems showing infrared luminosities comparable to those of local (ultra)luminous infrared galaxies. In \citet{baron18} and \citet{baron20} we used optical IFUs to spatially-resolve the stars and gas in two such galaxies. In this work, we used MUSE/VLT observations of 5 additional galaxies to study the spatial distribution of the stars and multiphased gas, and in particular, constrain the properties of the multiphased outflows. Our results and their broader implications are summarized below.

\textbf{(I) Tidal interactions and/or mergers.} All the 7 galaxies from our combined IFU sample show signatures of interaction or merger in their stellar or gas morphology or kinematics. In addition, 5 out of 7 galaxies show infrared luminosities of $11 < \log L_{\mathrm{SF}} / L_{\odot} < 12$. The galaxies in our sample are at different stages of interaction, including a pair of interacting galaxies at a distance of $\sim$50 kpc from each other, ongoing mergers with two visible nuclei at a distance $<5$ kpc, and a galaxy with no visible companion but with a tidal tail extending to a distance of 30 kpc. Interestingly, 4 out of our combined sample of 7 are at an early stage of the interaction, with visible companions at distances of $>$10 kpc. This suggest that post-starburst signatures in optical (strong H$\delta$ absorption) are not necessarily associated with post-merger systems. The observations are consistent with the idea that these galaxies have already had their first close passage, which led to the elevated SFR seen in infrared wavelengths and the post-starburst signatures seen in optical. Importantly, our observations suggest that H$\delta$-strong galaxies selected to have signatures of AGN and ionized outflows are more likely interacting starburst galaxies, rather than post-merger post-starburst galaxies. 

\textbf{(II) Broad kinematic components in optical emission lines do not necessarily trace outflows.} Using the MUSE observations, we performed a detailed analysis of the morphology, kinematics, flux distribution, and reddening of the broad kinematic components in each of the galaxies. Contrary to our initial expectation, the observations are consistent with nuclear (AGN or supernovae-driven) outflows only in 2 out of the 5 galaxies (4 out of the 7 galaxies in the combined sample). For some of the galaxies, the observations are more consistent with galactic-scale motions induced by the interaction/merger, a process that is often overlooked in studies of outflows. It is possible that our selection of galaxies with broad components in their optical emission lines favors interacting systems. This has significant implications for studies of ionized outflows in active galaxies, where it is a common practice to select systems with broader kinematic components in H$\alpha$ or [OIII] and classify the broad component as originating from an AGN-driven outflow. This poses a particular challenge for studies of higher redshift quasars, where spatially-resolved information is not available, and since quasar activity is linked to mergers in the local universe. Our results question the common assumption that broad kinematic components in the ionized emission lines trace primarily galactic-scale AGN or supernova-driven outflows. 

\textbf{(III) NaID emission and absorption are effective outflow tracers.} We detect NaID absorption in 3 out of the 5 galaxies. We detect a combination of blueshifted NaID absorption and redshifted NaID emission (classical P-Cygni profile) in two systems, which are also the only two systems whose ionized lines are consistent with a nuclear (AGN or supernova-driven) outflow. Contrary to the ionized emission lines, where it is not clear whether the blue/red wings trace inflows or outflows, the NaID P-Cygni profile does not suffer from this degeneracy. In case of an outflow, we expect to find blueshifted NaID absorption and redshifted NaID emission, while in the case of an inflow, the profile will be reversed with redshifted absorption and blueshifted emission. The blueshifted NaID absorption and redshifted NaID emission tend to trace separate regions within the galaxies, suggesting that, when detected, the NaID emission should be taken into account when estimating the mass and energetics of neutral outflows. We estimated the mass of the outflow, mass outflow rate, and kinetic power of the neutral gas that is traced by the NaID emission, and found them to be comparable to those derived from the absorption. We did not find a significant connection between the neutral and ionized outflows, but generally find $L_{\mathrm{NaID}} \sim 0.1 L_{\mathrm{H\alpha}}$.

\textbf{(IV) Properties of multiphased outflows.} For the two galaxies where the observations are consistent with an ionized nuclear (AGN or supernova-driven) outflow, we found mass outflow rates of 10 and 3 $M_{\odot}$/yr. These two galaxies also show a combination of NaID emission and absorption, with a total mass outflow rate of 7 and 21 $M_{\odot}$/yr. Assuming that the multiphased outflows are driven solely by the AGN, the wind coupling efficiency is estimated to be $\sim$0.5\%. However, both of these systems are infrared luminous galaxies ($\log L_{\mathrm{SF}}$ of 11.3 and 11.5 $L_{\odot}$), where we expect some contribution from supernovae to the observed winds. Therefore, the reported coupling efficiencies are upper limits.

The present study, together with the earlier papers published by our group (\citealt{baron22, baron23}) highlight the importance of using IFU observations combined with FIR-based estimates of SFRs in studies of galactic outflows. The IFU observations can be used to distinguish between different types of flows in galaxies (inflows versus outflows), different types of ionization mechanisms (AGN and SF ionization, shock excitation), and to map the kinematics and morphologies of neutral and ionized gas clouds. To form a more complete picture of the flows in these transitioning galaxies, it is also necessary to study the cold and warm molecular phases, traced by mm Carbon Monoxide lines and infrared H$_{2}$ lines respectively. We are currently involved in follow-up observations to trace these phases, and results will be reported in a forthcoming publication. 
 
\acknowledgments{
D. Baron is supported by the Carnegie-Princeton fellowship.
}

\software{Astropy \citep{astropy13, astropy18, astropy22},
		  IPython \citep{perez07},
          scikit-learn \citep{pedregosa11}, 
          SciPy \citep{scipy01},
		  matplotlib \citep{hunter07}}

\bibliography{ref_muse_sample}

\onecolumngrid

\appendix

\section{Voronoi binning parameters}\label{a:voronoi_bins}	

\floattable
\begin{deluxetable}{c c c c c c c c c}
\tablecaption{Voronoi binning parameters and resulting bins.\label{tab:voronoi_bins_parameters}}
\tablecolumns{9}
\tablenum{6}
\tablewidth{0pt}
\tablehead{
\colhead{Object ID} & \multicolumn{4}{c}{stellar population} & \multicolumn{4}{c}{emission and absorption lines} \\
\colhead{}          &  \colhead{minimum SNR} & \colhead{target SNR} & \colhead{N spaxels} & \colhead{N bins} & \colhead{minimum SNR} & \colhead{target SNR} & \colhead{N spaxels} & \colhead{N bins} \\
\colhead{(1)}       & \colhead{(2)} & \colhead{(3)} & \colhead{(4)} & \colhead{(5)} & \colhead{(6)} & \colhead{(7)} & \colhead{(8)}  & \colhead{(9)} \\
}
\startdata
J022912   & 10 & 250 & 11299 & 487 & 6   & 100 & 1974 & 137 \\
J080427   & 10 & 250 & 3719  & 187 & 6   &  50 & 337  &  96 \\
J112023   & 10 & 250 & 5027  & 545 & 8   & 100 & 1351 & 204 \\
J020022   & 10 & 250 & 4221  & 206 & 6   &  50 & 300  &  83 \\
J111943   & 10 & 250 & 5591  & 345 & 5   &  30 & 369  &  59 \\
\enddata
\tablecomments{\textbf{Columns.} (1): Object identifier. (2) and (6): minimum SNR below which a spaxel is masked out, for the stellar and emission/absorption line binning respectively. (3) and (7): target SNR given as an input to {\sc vorbin}. Spaxels will be binned together until the target SNR is reached. (4) and (8): the number of spaxels given as an inout to {\sc vorbin}. (5) and (9): number of bins given as an output by {\sc vorbin}.}
\end{deluxetable}

\section{Properties of the broad emission lines}\label{a:ionized_outflow_props}

\begin{figure*}
	\centering
\includegraphics[width=1\textwidth]{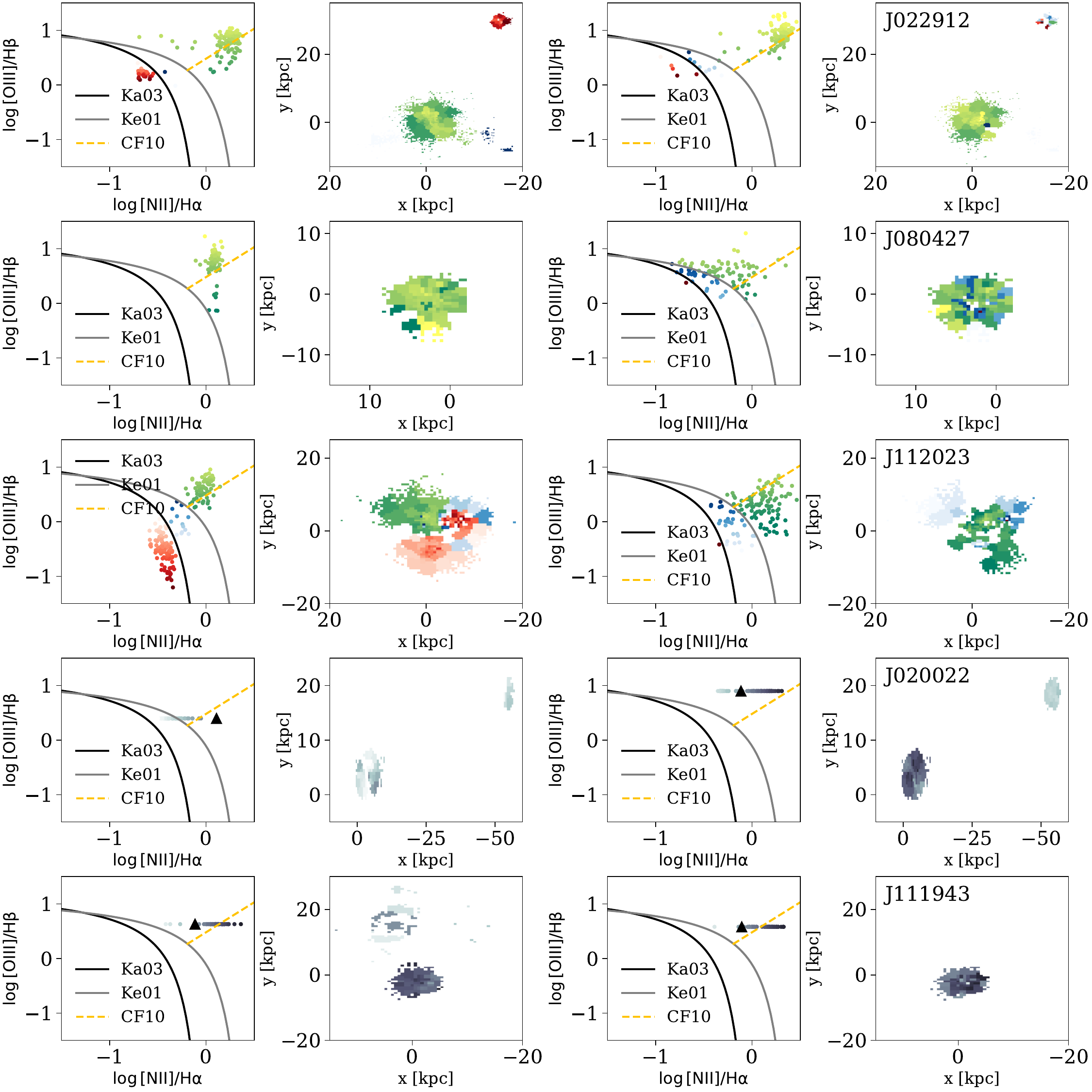}
\caption{\textbf{Line diagnostic diagrams for the ionized gas}. Each row represents a different galaxy. The first two columns show the narrow kinematic component, and the last two show the wings of the broad kinematic component. The first panel shows the location of different spaxels on the line diagnostic diagram [NII]/H$\alpha$ versus [OIII]/H$\beta$ (\citealt{baldwin81, veilleux87}). We mark the two separating criteria that are used to separate star forming from AGN-dominated galaxies (\citealt{kewley01, kauff03a}; Ke01 and Ka03 respectively), and the LINER-Seyfert separating line from \citet[CF10]{cidfernandes10}. The second panel shows the spatial distribution of the classification. For J020022 and J111943, the [OIII] is masked-out by the reduction pipeline since it coincides with the WFM-AO laser’s sodium lines. We therefore use the SDSS [OIII]/H$\beta$ and show the variation of the [NII]/H$\alpha$ ratio. }\label{f:BPT_narrow_and_wings}
\end{figure*}

\begin{figure*}
	\centering
\includegraphics[width=0.95\textwidth]{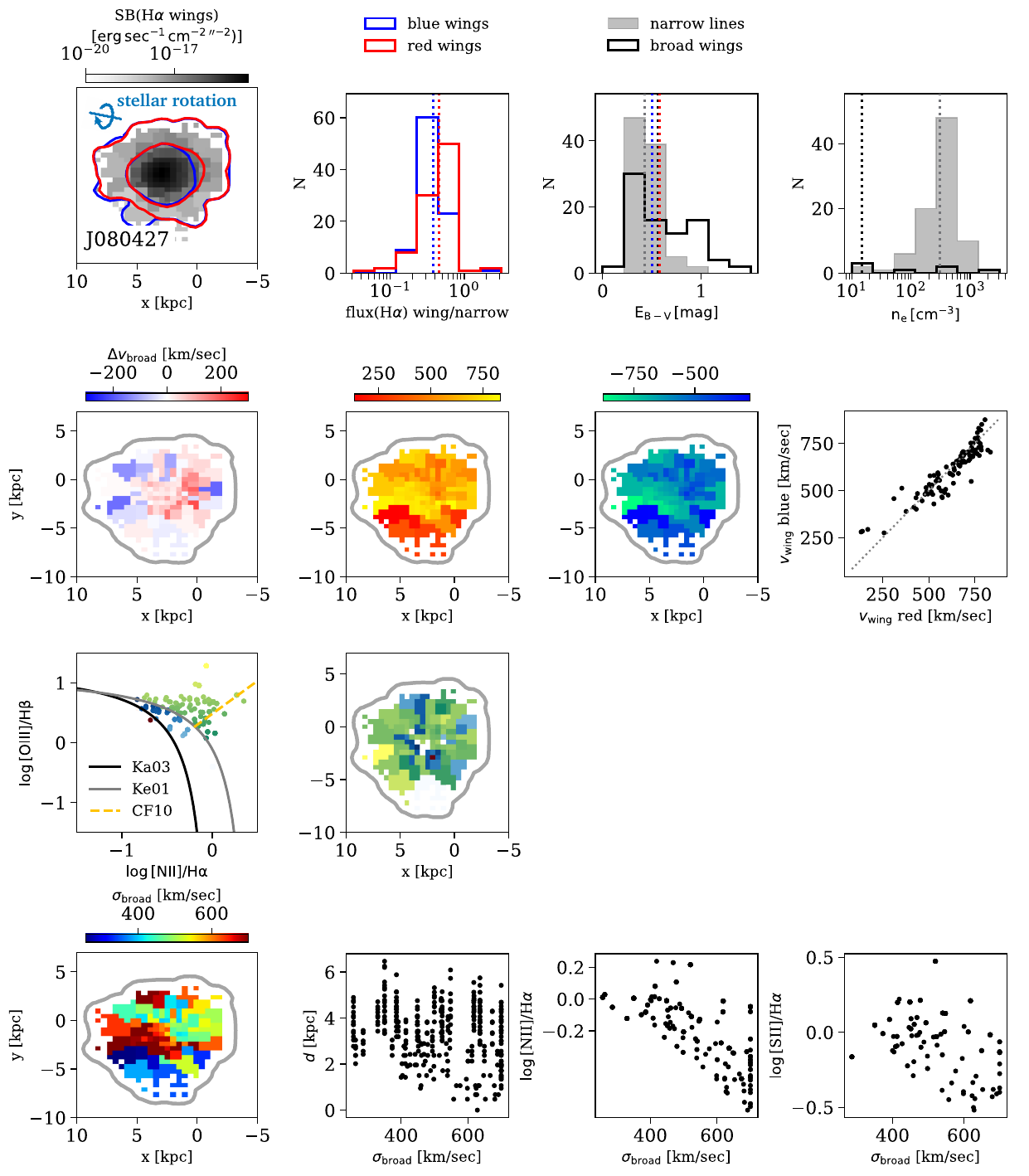}
\caption{\textbf{Broad line diagnostic diagrams for J080427.} Top row from left to right: (i) surface brightness of the broad H$\alpha$ wings (grey), with red and blue wings shown separately with contours, (ii) distribution of the flux in the wings, (iii) distribution of $\mathrm{E}_{B-V}$ measured towards the narrow lines (grey), broad wings (black), and red and blue wings, with dashed lines showing median values, and (iv) distribution of the electron density. Second row from left to right: (i) velocity of the broad kinematic component with respect to systemic, (ii) + (iii) maximum outflow velocity in the red and blue wings, and (iv) a comparison between the two. Third row: (i) the location of different spaxels on the line diagnostic diagram [NII]/H$\alpha$ versus [OIII]/H$\beta$ (\citealt{baldwin81, veilleux87}; separating criteria: \citealt{kewley01, kauff03a}; Ke01 and Ka03 respectively), (ii) and spatial distribution of the classification. Forth row: (i) velocity dispersion of the broad kinematic component, (ii) distance of each spaxel versus the velocity dispersion, and (iii) + (iv) the [NII]/H$\alpha$ and [SII]/H$\alpha$ broad line ratios versus the velocity dispersion.}\label{f:J080427_summary_plot}
\end{figure*}

\begin{figure*}
	\centering
\includegraphics[width=0.95\textwidth]{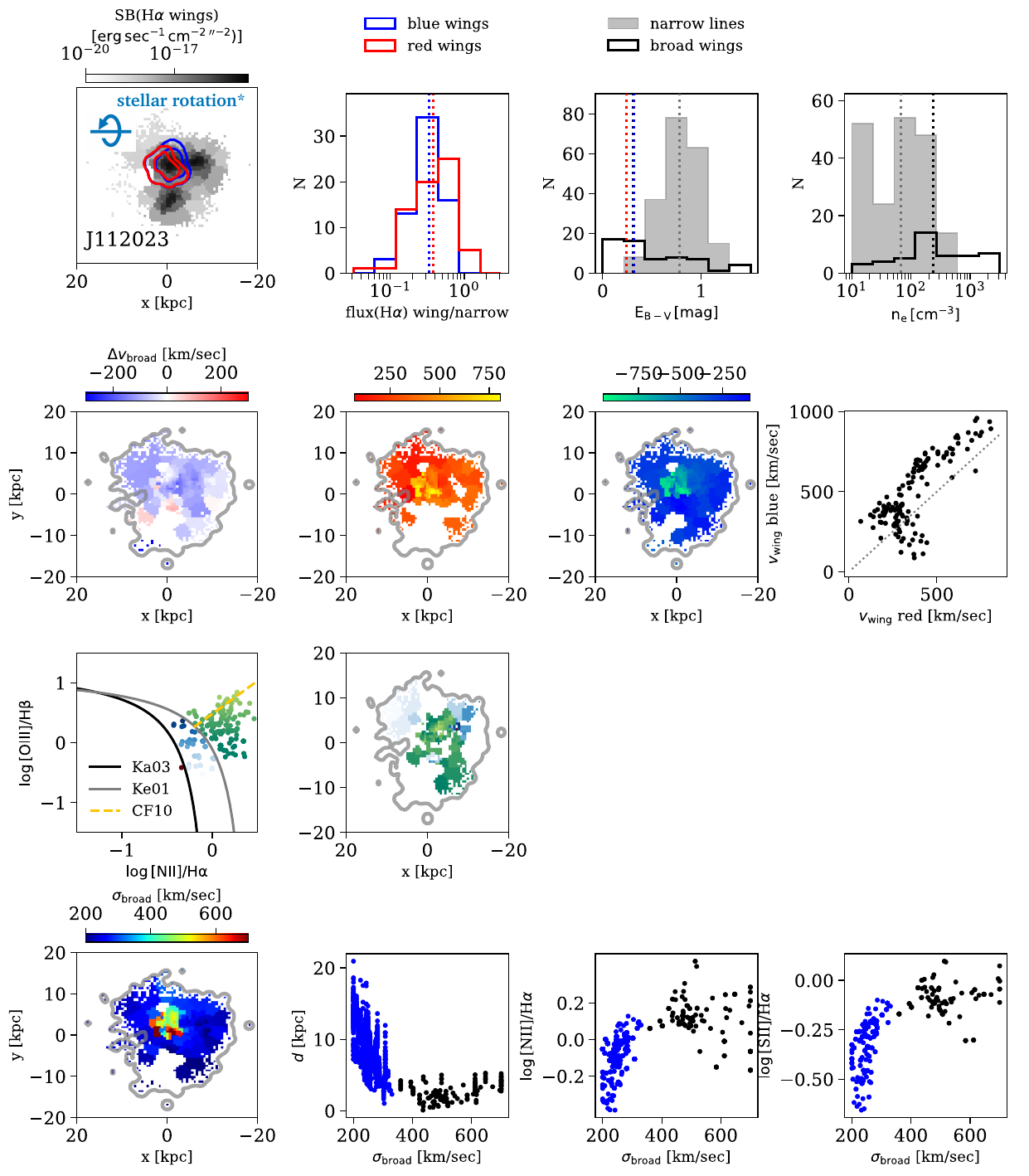}
\caption{\textbf{Broad line diagnostic diagrams for J112023.} See caption of figure \ref{f:J080427_summary_plot} for details about the panels. The stellar rotation axis marked in the leftmost panel in the first row is based on the stellar velocities measured within $\sim$7 kpc of the primary galaxy. Outside this region, J112023 shows disturbed stellar kinematics. }\label{f:J112023_summary_plot}
\end{figure*}

\begin{figure*}
	\centering
\includegraphics[width=0.95\textwidth]{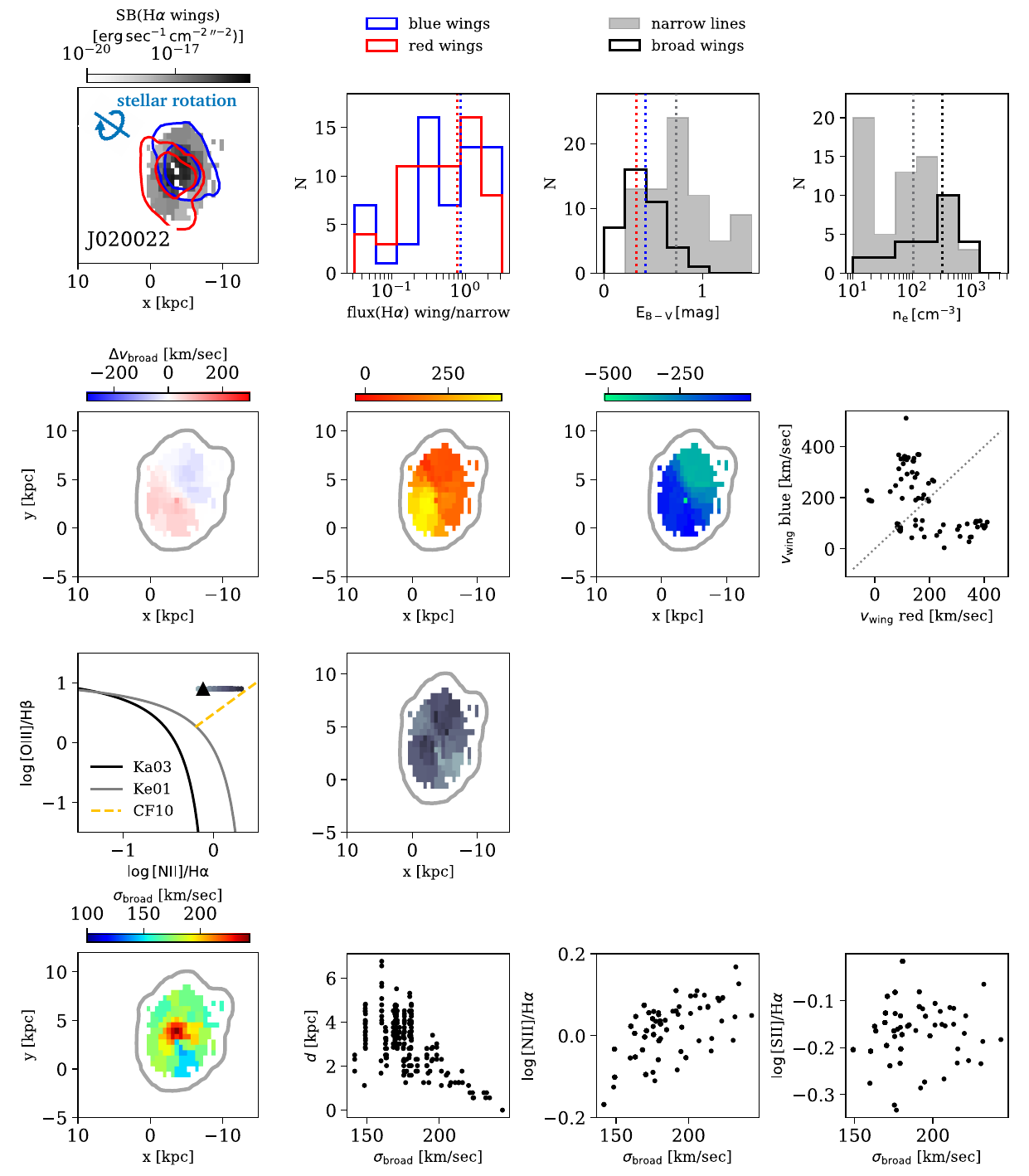}
\caption{\textbf{Broad line diagnostic diagrams for J020022.} See caption of figure \ref{f:J080427_summary_plot} for details about the panels. For J020022 and J111943, the [OIII] is masked-out by the reduction pipeline since it coincides with the WFM-AO laser’s sodium lines. We therefore use the SDSS [OIII]/H$\beta$ and show the variation of the [NII]/H$\alpha$ ratio. }\label{f:J020022_summary_plot}
\end{figure*}

\begin{figure*}
	\centering
\includegraphics[width=0.95\textwidth]{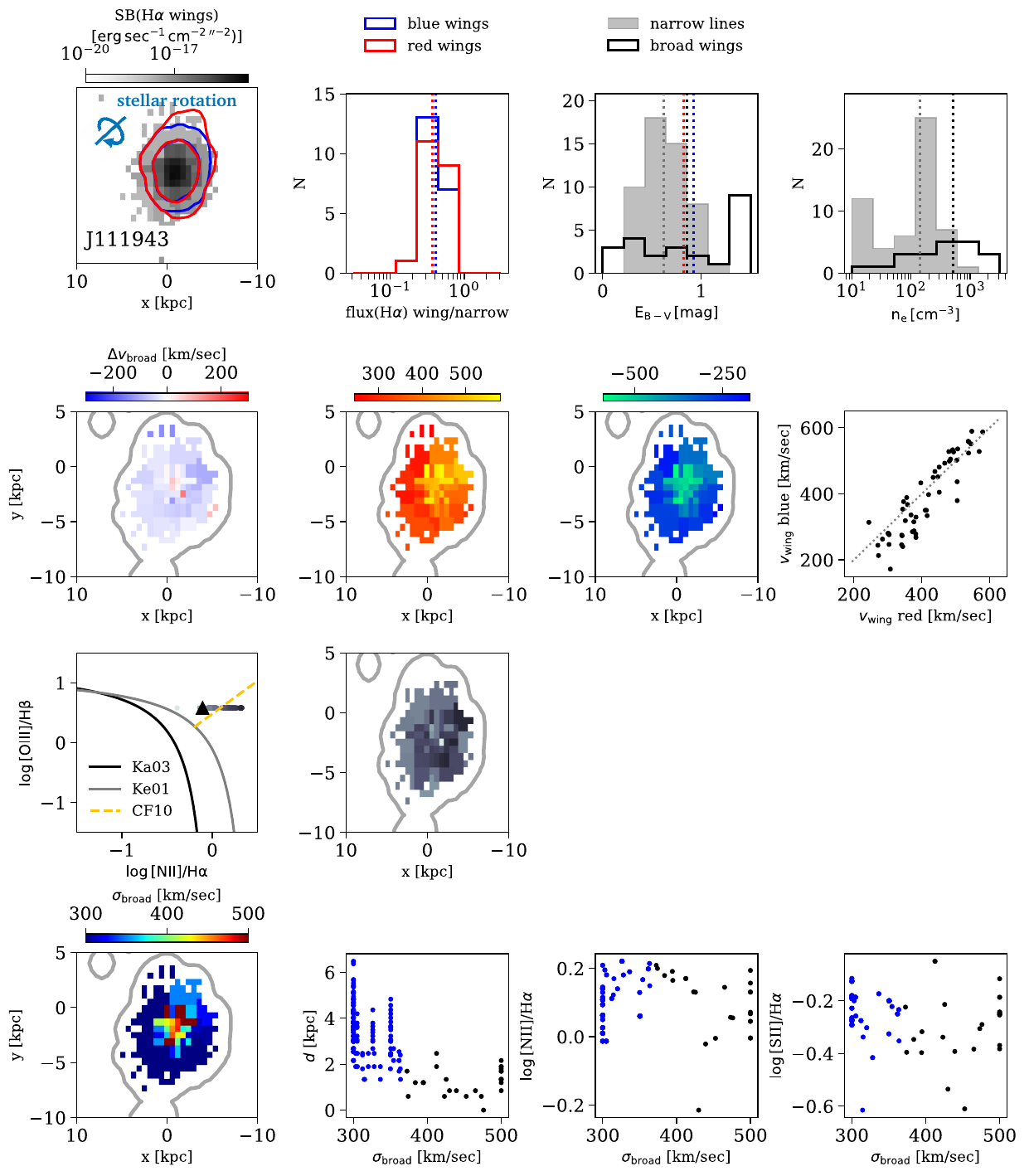}
\caption{\textbf{Broad line diagnostic diagrams for J111943.} See caption of figure \ref{f:J080427_summary_plot} for details about the panels. For J020022 and J111943, the [OIII] is masked-out by the reduction pipeline since it coincides with the WFM-AO laser’s sodium lines. We therefore use the SDSS [OIII]/H$\beta$ and show the variation of the [NII]/H$\alpha$ ratio. }\label{f:J111943_summary_plot}
\end{figure*}

\section{NaID profile properties}\label{a:NaID_fitting}

\begin{figure*}
	\centering
\includegraphics[width=0.85\textwidth]{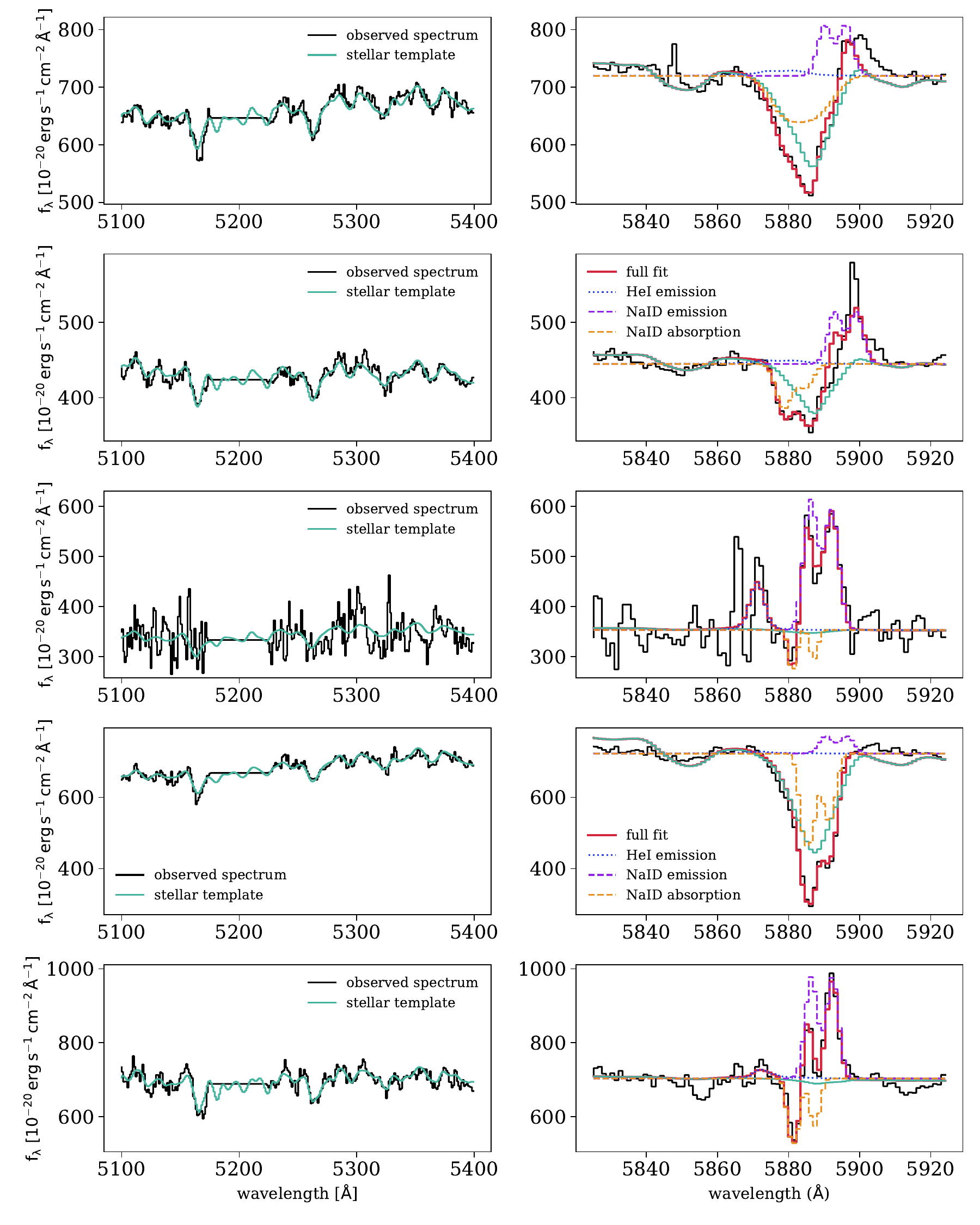}
\caption{\textbf{Examples of NaID profile fitting for J022912.} Each row represents a different binned spectrum. The left column shows the observed flux (black) and the best-fitting stellar model by {\sc ppxf} around the stellar MgIb absorption complex. The good correspondence between the observed spectrum and the stellar model suggests that the excess NaID absorption/emission is not due to a bad stellar fit nor of a stellar origin. The right column shows the observed flux (black) around the NaID region, along with the best-fitting model. The full fit is shown in red. The separate model components are also shown: HeI emission (dotted blue), NaID emission (dashed purple), and NaID absorption (dashed orange). The full fit also depends on the best-fitting stellar model, shown in green. 
}\label{f:NaID_fitting_example}
\end{figure*}

\begin{figure*}
\includegraphics[width=0.9\textwidth]{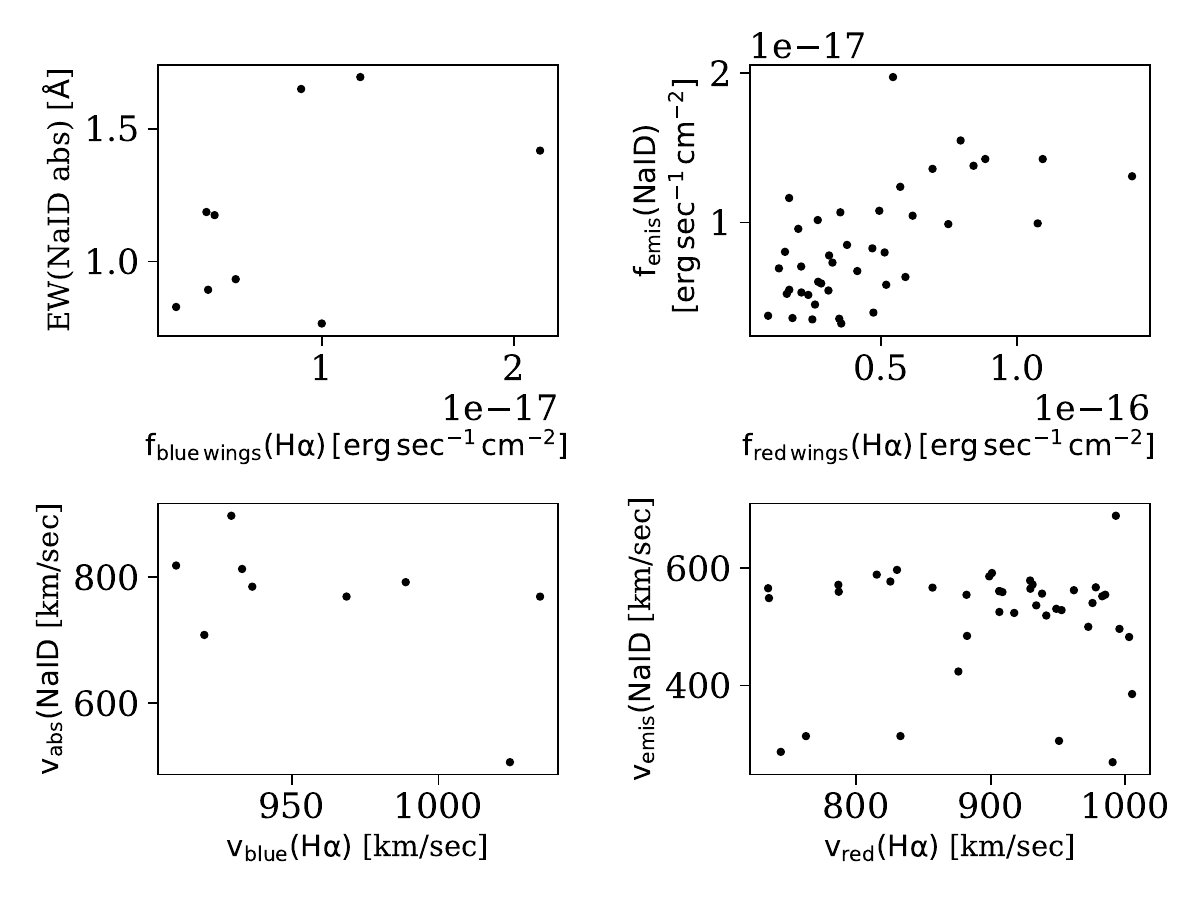}
\caption{\textbf{Comparison between the ionized and neutral outflow of J022912.} The top row compares the blueshifted absorption EW(NaID) to the blue wing of the broad H$\alpha$ emission (left) and the redshifted NaID flux to the red wing of the broad H$\alpha$ emission (right). The bottom panel compares the derived outflow velocities, where the left panel compares the blueshifted NaID absorption velocity to the velocity of the blue H$\alpha$ wing, and the right panel compares the redshifted NaID emission velocity to the velocity of the red H$\alpha$ wing.} 
\label{f:J022912_neutral_to_ionized_relation}
\end{figure*} 

\begin{figure*}
\includegraphics[width=0.9\textwidth]{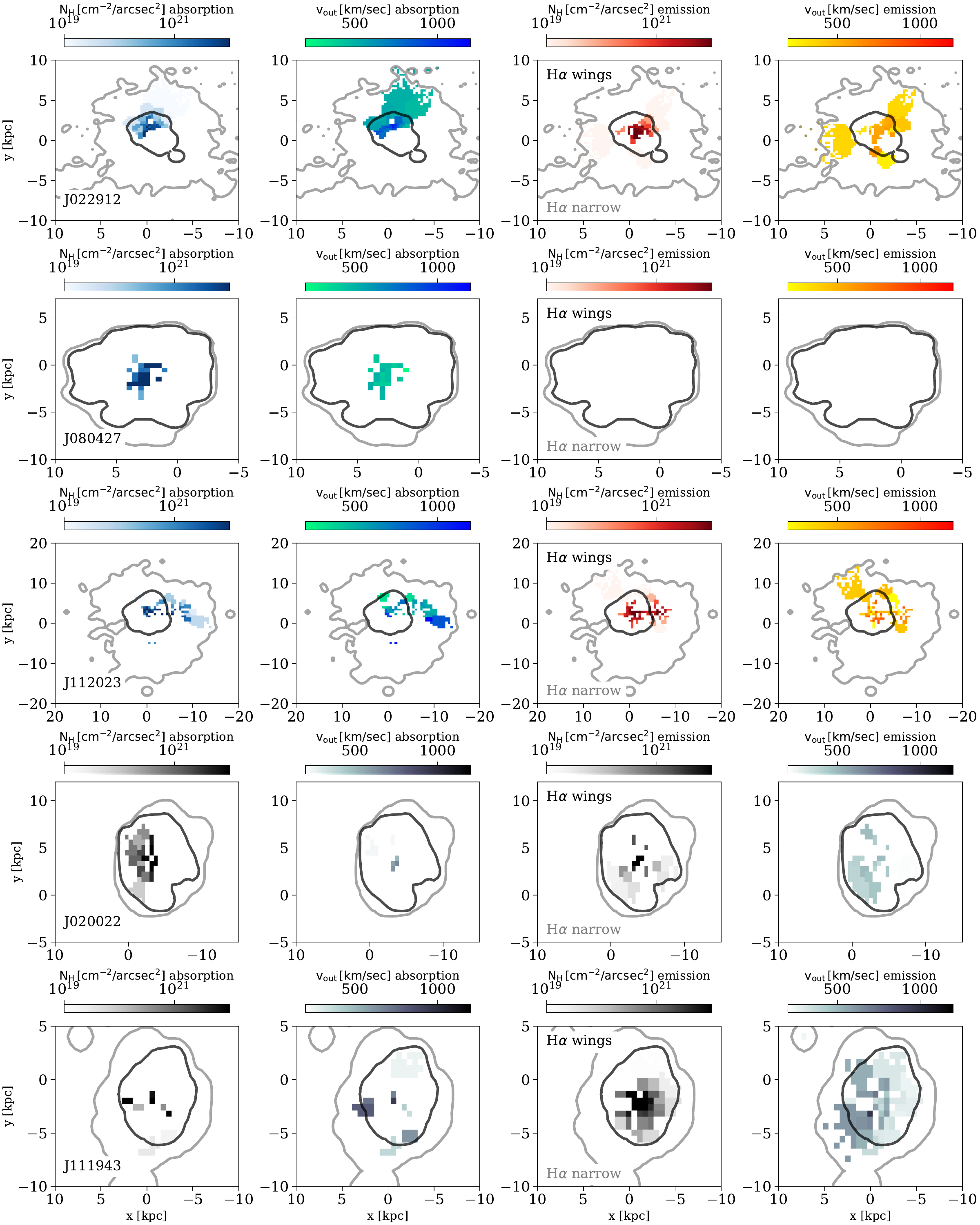}
\caption{\textbf{Properties of the neutral outflow.} Each row represents a different galaxy. The first two columns show the Hydrogen column density and the adopted velocity of the outflow traced by blueshifted NaID absorption. The second two columns show the Hydrogen column density and the adopted velocity of the outflow traced by the redshifted NaID emission. The grey contours represent the extent of the narrow line-emitting ionized gas, and the black represent the ionized outflow. For J020022 and J111943, neither NaID absorption nor emission is detected, and show the upper limits on these properties. } 
\label{f:neutral_outflow_properties}
\end{figure*} 

\end{document}